\documentclass[11pt]{article}
\usepackage{amsmath}
\usepackage{amsfonts}
\usepackage{amssymb}
\usepackage{amsthm}
\usepackage{tikz}
\usepackage{tikz-cd}
\usepackage{mathtools}
\usepackage{graphicx}
\usepackage[letterpaper,hmargin=1in,vmargin=1in]{geometry}
\usepackage[usenames,dvipsnames]{pstricks}
\usepackage{epsfig}
\usepackage{pst-grad} 
\usepackage{pst-plot} 
\usepackage[space]{grffile} 
\usepackage{etoolbox} 
\usepackage{blindtext,enotez}
\usepackage{cite}
\usepackage[numbers]{natbib}
\makeatletter 
\patchcmd\Gread@eps{\@inputcheck#1 }{\@inputcheck"#1"\relax}{}{}
\makeatother 
\let\footnote\endnote

\usepackage{ifthen}

\newcommand{\imat}{{\mathrm{i}}}

\newcommand{\DEF}{\stackrel{\mathrm{def}}{=}}
\newcommand{\DEFt}{\;\smash{\stackrel{\mbox{\tiny def}}{=}}}

\usepackage{hyperref}
\hypersetup{pdfpagemode=UseNone}

\usepackage{amsfonts}
\usepackage{amssymb}
\usepackage{amsmath}
\usepackage{latexsym}
\usepackage{amsthm}
\usepackage{color}
\usepackage{enumerate}
\usepackage{caption}

\begin{document}

\begin{flushleft}
{\huge
\textbf\newline{Riemann surfaces of complex classical trajectories and tunnelling splitting in one-dimensional systems}
}
\newline
\\
{\bf Hiromitsu Harada}\textsuperscript{1},
{\bf Amaury Mouchet}\textsuperscript{2},
{\bf Akira Shudo}\textsuperscript{1}
\\
\bigskip
{\footnotesize
{1} Department\ of\ Physics,\ Tokyo\ Metropolitan\ University, Minami-Osawa, Hachioji, Tokyo 192-0397, Japan
\\
{2} Laboratoire de Math${\rm \acute{e}}$matiques et de Physique Th${\rm \acute{e}}$orique, Universit${\rm \acute{e}}$ Fran{\c c}ois Rabelais de Tours-CNRS (UMR 7350), F\'ed\'eration Denis Poisson, Parc de Grandmont 37200 Tours, France
\\
\bigskip
* harada-hiromitsu@ed.tmu.ac.jp, Amaury.Mouchet@lmpt.univ-tours.fr, shudo@tmu.ac.jp
}
\end{flushleft}

\begin{abstract}
The topology of complex classical paths is investigated to discuss
quantum tunnelling splittings in one-dimensional systems. Here the
Hamiltonian is assumed to be given as polynomial functions, so the
fundamental group for the Riemann surface provides complete
information on the topology of complex paths, which allows us to
enumerate all the possible candidates contributing to the
semiclassical sum formula for tunnelling splittings.  This naturally
leads to action relations among classically disjoined regions,
revealing entirely non-local nature in the quantization condition. The
importance of the proper treatment of Stokes phenomena is also
discussed in Hamiltonians in the normal form.
\end{abstract}

\section{Introduction}
By definition, tunnelling is a purely quantum effect that cannot be
described by any real solution of the classical dynamics.  One of the
best known signature of it is provided by the splittings in the energy
spectrum of a quantum one-dimensional particle in a symmetric
double-well potential. A state localised in one well is coupled to its
parity-related twin localised in the other well to form a
symmetric/antisymmetric doublet of eigenstates delocalised in both
wells whose energies differ by a small amount that depends
exponentially on the inverse of the Planck constant~$\hbar$ or on any
classical parameter. Even though no classical real solution connects
the two wells, by extending classical dynamics from real to complex
plane and applying the WKB method, one can actually capture such
nonclassical phenomena.  {\it Instanton} is broadly recognized as a
classical path running in the complex plane, which has capability of
describing tunnelling in the double-well potential or degenerated
vacua in the fields theory \cite{Coleman}.  The instanton was
originally obtained by performing the so-called Wick rotation of
time~$t\to\imat t$.  More generally, one may find in references
\cite{Freed,GeorgeMiller,Miller,Weiss,Carlitz,Ilgenfritz,CreaghWhelan}
some applications of complexifying time in different contexts but the
arguments and techniques developed there have mainly been made to
understand quantum tunnelling in one dimension.

On the other hand, quantum tunnelling has received renewed interest
for these two decades.  One driving force for this is that our
understanding for classical dynamics has been proceeded considerably
and we recognized that qualitative and essential differences in nature
of classical dynamics underlie between one and multi-dimensions.  In
particular, multidimensional systems are known to be nonintegrable in
general, which naturally leads to pay attention on the nature of
quantum tunnelling in chaotic situations
\cite{Bohigas93,Creagh98,KS11}.\endnote{We should also mention that there is an abundant literature on these
matters in mathematical physics. Two lines of thought may be identified
(i) the first one, initiated by Harrell \cite{Harrell78a} for the
quartic 1d-potential, relies on instantons, that were later seen as
geodesics for a measure, called the Agmon measure, that evaluates the
quantity of ``classical interdiction''.  The extension to
multidimensional systems was made by Davies \cite[Th.~4]{Davies82a}
(exponential majoration), then by Simon \cite[Th.~1]{Simon83b}
(exponential behaviour), Helffer \& Sj\"ostrand
\cite{Helffer/Sjostrand84a,Helffer/Sjostrand86a,Helffer/Sjostrand87a}
(prefactor of the exponential). (ii) The second school plays also with
the exponential behaviour of the wavefunctions outside the wells but
by directly complexifying the time in the Schr\"odinger equation and
then applying the semiclassical methods developed by Maslov
\cite{Dobrokhotov+91a,Dobrokhotov/Kolokoltsov93a}. These two schools
are mainly concerned by the tunnelling doublet of lowest energy
(see however \cite{Martinez88a}) and surprisingly enough the
authors seem to consider as secondary the crucial hypothesis that
allows to explain the dichotomy between tunnelling in integrable and
non-integrable systems.  This hypothesis is let implicit in
\cite{Simon83b} (one has to isolate the Agmon geodesics, see condition
(1.1) of Th. (1.1) in \cite{Martinez87a}, this hypothesis is not even
mentioned in Th.~1.5 of \cite{Simon84a} and appears indirectly in
\cite{Helffer/Sjostrand85a} through the hypothesis of the non
degeneracy of the instanton trajectories, see also \S~IV, hypothesis
$H_4$ in \cite{Sordoni97a}) and we were not even able to identify it
in the work of the Russian school.}

There are actually two tasks in
performing the semiclassical analysis. The first one concerns how to
establish a proper semiclassical formulation providing observed
quantities, such as tunnelling splittings.  Our second task is to find
or even enumerate the inputs ---expected to be real or complex
classical quantities--- which are necessary for the semiclassical
analysis.

Concerning formalisms in the semiclassical analysis, if we restrict
our interest to energy splittings invoked by quantum tunnelling,
explicit and closed formulas are rather limited, although energy
splittings are quantities in which tunnelling effects could typically
be observed even in experiments.  This is the case even in
one-dimensional situations \cite{Landau,Connor,Garg,instanton,JMS}.

The second task would also not be so easy because we need to be
thoroughly familiar with classical dynamics in the complex plane.  In
the case of discrete dynamical systems, fortunately enough, we could
make full use of the results gained in recent progress on
multidimensional complex dynamical systems and a close link between
signatures of quantum tunnelling and complex classical dynamics was
discovered \cite{Shudo1,Shudo2}.  On the other hand, for continuous
flow systems, our knowledge about the dynamics in the complex plane is
rather fragmental and not enough to reach a unified perspective.  Much
efforts have been made to explore the nature of singularities in the
complex time plane by studying simple scattering models closely
\cite{KTakahashi1,KTakahashi2,KTakahashi3}, but the analyses were not
exhaustive and remain rather heuristic.  This is mainly because the
models examined there were still not simple enough in the sense that
the nature of singularities appearing in the associated classical
dynamics remained too intricated to be handled in a rigorous manner.

Under such circumstances, the aim of the present paper is to focus on
the second issue and establish models which allow full enumeration of
complex orbits necessary for the semiclassical analysis of tunnelling
splittings.  This will be achieved for the systems whose Hamiltonians
are given as polynomial functions.  If Hamiltonian functions are
polynomial, any local classical quantities are algebraic functions of
the dynamical variables, which greatly simplifies the Riemann sheet
structure and makes it possible to develop rigorous arguments on
classical dynamics in the complex plane.  In particular, because there
is a finite number of algebraic singularities and no essential
singularities, we can easily describe the associated Riemann surface
and its fundamental group.

The organization of the present paper is as follows.  In section 2, we
introduce the semiclassical formula for tunnelling splittings which
was derived in Ref. \cite{instanton}.  Our argument will throughout be
based on it.  We also mention the limitation of our analysis,
especially in view of the Stokes phenomenon.  Section 3 is devoted to
explaining our strategy to enumerate topologically distinct complex
paths.  A key idea is to examine the fundamental group of the Riemann
surface for the associated function, which makes it possible to obtain
a complete list complex paths.  In sections 4 and 5, simple models,
one-dimensional systems with double- and triple-well potentials, are
recast with special focus on the method of listing the relevant
complex paths we introduced in section 3.  An advantage in taking
such an approach is that one can find non-trivial global relations
among action integrals appearing in the semiclassical formula.  In
section 6, we show that such action relations imply a sufficient
condition under which distinct potential wells are simultaneously
quantized.  Since the condition originates only from the global
topology of the Riemann surface, the argument applies even in
asymmetric multi-well potential systems.  In section 7, we apply our
fundamental-group-based inventory to a richer integrable model
constructed with the help of Hamiltonian normal forms.  However, in
section 8, we emphasize that handling of the Stokes phenomenon done in
the cases of double- and triple-well models are improper for the
normal form Hamiltonian model by showing a counterexample for which a
naive prescription in dealing with the Stokes phenomenon does not
work.

\section{Semiclassical Formula for the Tunnelling splitting}

In this section, we introduce a semiclassical formula on which we will
rely throughout the following analysis for tunnelling splitting in
multi-well potential and normal form Hamiltonian systems.  In
Ref. \cite{instanton}, a semiclassical trace formula for tunnelling
splittings has been derived and it was shown to work well in
predicting tunnelling splittings for a symmetric triple-well potential
system.  Below we briefly explain the formula to show how {\it complex
  classical orbits} come into play in determining tunnelling
splittings (see more details in Appendix A and Ref. \cite{instanton}).

Let us consider a one-dimensional constant classical Hamiltonian
$H(p,q)$ having reflection symmetry with respect to the canonical
variables $p$ and $q$:
\begin{eqnarray}\label{eq:Hparity}
H(p,q)=H(-p,-q). 
\label{syH}
\end{eqnarray}
The energies~$E^{\pm}_{n}$ and the associated
eigenstates~$|\phi^{\pm}_{n}\rangle$ of the corresponding quantum
model are given by
\begin{eqnarray}
\hat H|\phi^{\pm}_{n}\rangle=E^{\pm}_{n}|\phi^{\pm}_{n}\rangle, 
\end{eqnarray}
where~$\hat H\DEF H(\hat p, \hat q)$ with $\hat p$ and $\hat q$ being
the canonical operators associated with~$p$ and~$q$, respectively.  The
superscripts $\pm$ stand for the symmetric/antisymmetric states and
tunnelling manifests itself through the splittings $\Delta E_{n}\DEFt
E^{-}_{n}-E^{+}_{n}$.

In Ref. \cite{instanton}, a semiclassical formula for the energy
splitting $\Delta E_{n}$ has been derived as
\begin{eqnarray}
\Delta E_{n}\sim\frac{\hbar}{2T}\sum_{{\rm cl}} (-1)^{\mu+1}e^{\imat S_{{\rm cl}}/\hbar},
\label{splitting}
\end{eqnarray}
where the sum is taken over all the classical paths with energy $E\sim E_{n}^{\pm}$ 
such that $q(T)=-q(0)$ and $p(T)=-p(0)$ for a given time
interval $T$.  Although the time interval $T$ appears explicitly in
the formula, it has been shown in Ref. \cite{instanton} that the
right-hand side of~(\ref{splitting}) becomes independent of~$T$ as
long as Im\,$T$ is taken to be large enough compared to the typical
(real) period of the classical system.

The quantities $S_{{\rm cl}}$ and $\mu$ denote the classical action
of the path $\Gamma$
\begin{eqnarray}
\label{classical_action}
S_{\rm cl} = \int_{\Gamma} p(q) dq, 
\end{eqnarray}
and the Maslov index \cite{Maslov}, respectively. The function $p(q)$ is defined by
\begin{equation}\label{eq:HpqE}
  H(p,q)=E,
\end{equation}
where~$E\simeq E^\pm_n$ and we will always left implicit the
dependence on~$E$.  Since there exist no real classical paths
connecting two classically disjointed regions, the path $\Gamma$ runs
in the complex plane.

Formula (\ref{splitting}) comes from the saddle point approximation, therefore, in order to apply it, two steps can be identified. The
first one is to list all the possible complex paths that could
contribute to the sum in the formula and the second step is to select
in this list of candidates those that actually contribute to~$\Delta
E_n$. As far as the first step is concerned, in general, even in the
simplest models such as double-well potential systems,
the classical solutions with appropriate boundary conditions
occur in families of infinite numbers and it is a non-trivial task to
enumerate all these stationary paths.  Even after enumerating all the
possible candidates, it is known that not all of them do not
necessarily remain as final contributions.  This is because the Stokes
phenomenon occurs in the complex plane, and some saddles have to be
excluded from the final contribution. The second step we should
consider is therefore to find a proper way of handling the Stokes
phenomenon.

In Ref.~\cite{instanton}, the formula (\ref{splitting}) was
satisfactorily tested in some standard models with a procedure for
achieving the first step that does not guarantee that all the possible
complex stationary path were considered.  To justify the adopted
method and to have a better control on the approximations, we need to
establish a systematic way, based on more rigorous grounds, to achieve
step one.  Our subsequent argument will be focused mainly on this
step.  Concerning the second step, although the Stokes phenomenon
could be now captured as a well recognized object
\cite{Voros83,DDP97,KT06} and even within the scope of rigorous
arguments thanks to recent progress of the so-called {\it exact WKB
  analysis}, or {\it resurgent theory}\cite{KT_Book}, we will not take
into account the Stokes phenomenon based on such recent developments,
rather treat the Stokes phenomenon in a heuristic way, as explained
below.

As is easily seen, classical actions associated with complex paths
have imaginary parts, and the complex path(s) with the most dominant
weight are supposed to have minimal imaginary action. Note that such
an argument of course holds only after handling the Stokes phenomenon
in an appropriate manner.  Our task here is therefore to enumerate all
the possible candidate complex paths and then to specify the complex
path with the smallest imaginary action out of the candidates.

The strategy for the first step is to examine the {\it fundamental
  group} of the Riemann surface $R$ of the function $p(q)$ since the
fundamental group provides the topological independent paths on a given surface.  In addition to
such information we also need to specify singularities of the function
$p(q)$. This is because, by virtue of Cauchy theorem, the value of the
classical action~(\ref{classical_action}) is affected when a
continuous deformation of~$\Gamma$ crosses singularities.

\section{Fundamental group of Riemann surfaces of algebraic functions}

In this section, we show how the fundamental group of the Riemann
surface $R$ of the function $p(q)$ helps to construct the
path~$\Gamma$ along which the classical
action~(\ref{classical_action}) is computed.  In what follows, we
assume that our Hamiltonian $H(p,q)$ is expressed as a polynomial
function of $p$ and $q$ and the polynomial is irreducible.  The former
condition allows to obtain the complete list of the complex paths
contributing to the semiclassical sum (\ref{splitting}) and the latter
condition ensures that the Riemann surface of the function $p(q)$ is
connected.

$H$ being a polynomial, the function~~$p(q)$ defined
by~(\ref{eq:HpqE}) is an algebraic function and therefore has at most
finitely many singularities \cite{math3} that are points where $p(q)$
has a pole or a branch point.  Since our Riemann surface $R$ is
constructed from an algebraic function and assumed to be irreducible,
it is homeomorphic to a surface of a finite genus $g$, or $g$-fold
torus for short, accompanied with finite number of holes associated
with singularities of the function under consideration.  The genus of
the surface is given by the formula $g= w/2- d+ 1$, where $w$ is the
ramification index and $d$ is the highest degree of $p$ in the
polynomial in question.  Especially, $w$ is equal to the number of
branch points if all branch points are square-root type, i.e., the
function is double-valued near each branch point. For example, in
multi-well potential systems discussed in sections 4 and 5, and the
normal form Hamiltonian system in section 7 as well, $p(q)$ is shown
to be double-valued functions near each branch point.

The fundamental group on the Riemann surface is introduced as the
group whose elements are identified through homotopy equivalence of
curves on the surface.  For the $g$-fold torus, there exist $2g$
independent homotopically equivalent loops, and following the
convention we call the half of them $\alpha_i$-loop and the rest
$\beta_i$-loop ($1\le i \le g)$.  The loops $\alpha_{i}$ and
$\beta_{i}$ are often called {\it homology basis} in the literature
\cite{math3}.

\begin{figure}[htbp]
\begin{center}
\includegraphics[width=12cm]{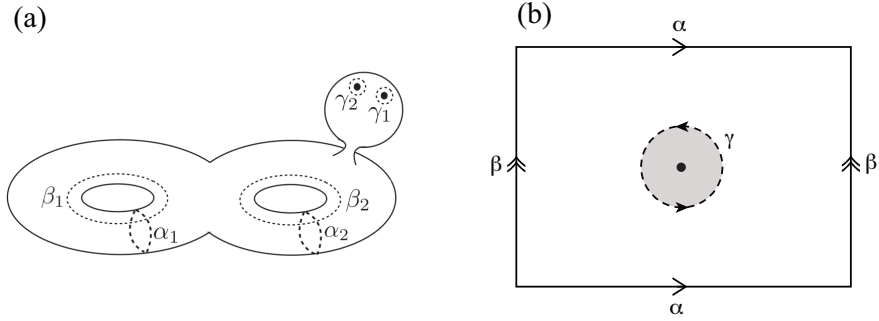}
\caption{
(a) An example of the Riemann surface. Here the case for the 2-fold torus with
two holes is presented.   
Homology bases of the fundamental group are shown
as $\alpha_{1},\alpha_{2},\beta_{1},\beta_{2},\gamma_{1}$ and $\gamma_{2}$.
(b) A graphical proof for the relation (\ref{eq:homrelations}). A simple torus with a hole is here assumed.
}
\label{gfoldtorus}
\end{center}
\end{figure}

When computing the action integral (\ref{classical_action}) one must include the
contribution of singularities which could provide non-zero residues,
when deforming~$\Gamma$.  This means that the associated
fundamental group should be replaced by the one incorporating
singularities of the function $p(q)$.  The Seifert-Van Kampen theorem
tells us that the fundamental group for a surface with holes is
obtained as the product of the fundamental group for the original
$g$-fold torus and that of a sphere with $m$ holes, where $m$ is the
number of holes \cite{Kosniowski}, which appear as either poles or
branch points in the present situation.  We call the loop encircling a
hole the $\gamma_i$ loop ($1\le i \le m)$, again following the
convention. The loop $\gamma_{i}$ here is taken to be a small closed
loop around each hole (see figure \ref{gfoldtorus}).

We note also from the Seifert-Van Kampen theorem 
that the elements $\alpha_{i},\beta_{i}$ and
$\gamma_{i}$ of the fundamental group satisfy a relation,
\begin{eqnarray}\label{eq:homrelations}
\prod_i\alpha_i\beta_i\alpha^{-1}_i\beta^{-1}_i=\prod_i\gamma_i
\end{eqnarray}
implying that all the loops $\alpha_{i},\beta_{i}$ and $\gamma_{i}$ are not independent with each other.
We hereafter assume that one of $\gamma_{i}$-loops, say $\gamma_m$, is expressed in terms of the other loops. 
We just graphically show in Fig. \ref{gfoldtorus} why the relation (\ref{eq:homrelations}) follows in the simplest situation where a simple torus with $g=1$ is connected with
 a sphere with a hole. 

Using the elements of the fundamental group, we can now enumerate all the topologically distinct paths obtained from a reference path 
$\Gamma_0$.  More concretely, for an arbitrarily chosen 
reference path $\Gamma_0$ with the fixed 
initial and final ends in the $q$-plane, topologically independent 
paths associated with the reference path $\Gamma_0$ are expressed as 
\begin{eqnarray}\label{general_expression}
\Gamma = \Gamma_0+\sum_{i=1}^{g}n_{\alpha_{i}}\alpha_{i}+\sum_{i=1}^{g}n_{\beta_{i}}\beta_{i}+\sum_{i=1}^{m-1}n_{\gamma_{i}}\gamma_{i},
\end{eqnarray}
where $n_{\alpha_{i}},n_{\beta_{i}}$ and $n_{\gamma_{i}}$ are integers and will be called winding numbers hereafter. 
In what follows we apply the scheme formulated in this way to a couple of concrete examples, some of them are the systems already well studied.

\section{Double-well potential case}
\begin{figure}[htbp]
\begin{center}
\includegraphics[width=7cm]{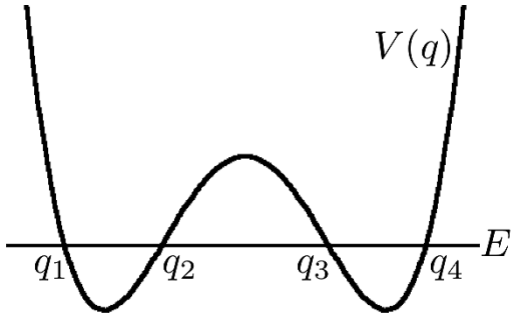}
\caption{The double-well potential $V(q)$.}
\label{doublewellpot}
\end{center}
\end{figure}

\begin{figure}[htbp]
\begin{center}
\includegraphics[width=12cm]{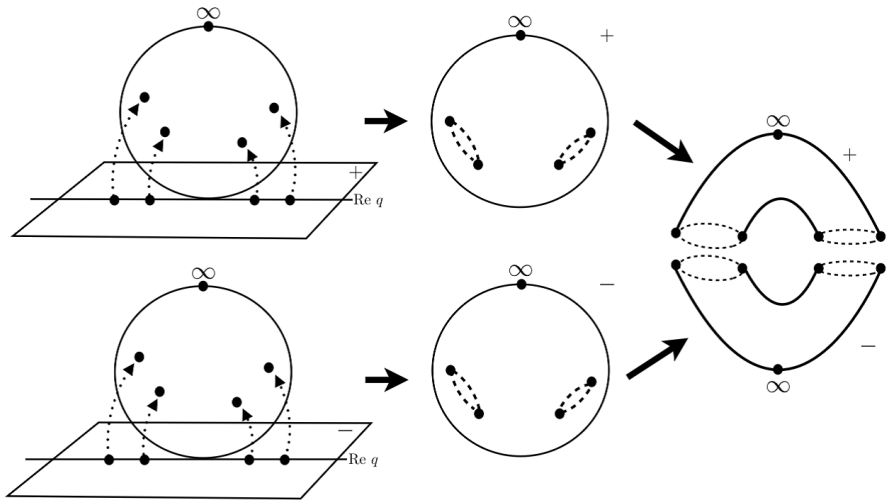}
\caption{Deformation of Riemann spheres to a torus is shown 
in the double-well potential case. The black dots and the dashed lines represent branch points and branch cuts, respectively. $\pm$ signs show the branches of $p(q)$.}
\label{fig2}
\end{center}
\end{figure}

As a simple example, we first discuss a double-well potential system:
\begin{eqnarray}
H(p,q)=\frac{p^2}{2}+V(q),
\end{eqnarray}
\begin{eqnarray}
V(q)=E+(q-q_1)(q-q_2)(q-q_3)(q-q_4).
\label{DWpotential}
\end{eqnarray}
Here $q_i ~( 1\le i \le 4)$ are real parameters satisfying $q_1<q_2<q_3<q_4$ 
and $E$ is the total energy. 
We further assume that the potential function is symmetric, that is $q_1=-q_4,q_2=-q_3$
(see figure \ref{doublewellpot}) in accordance with~(\ref{eq:Hparity}) even though this symmetry condition
is not relevant for topological considerations. 
From~(\ref{eq:HpqE}) we find 
\begin{eqnarray}
p(q)= \pm \sqrt{-2(q-q_1)(q-q_2)(q-q_3)(q-q_4)}.
\end{eqnarray}
The function $p(q)$ has 
four branch points at $q = q_i ~( 1\le i \le 4)$, which are all located on the real axis and one can choose
the intervals~$[q_1,q_2]$ and~$[q_3,q_4]$ as two cuts defining a Riemann surface with two leaves.
As shown in figure \ref{fig2},  we project each leaf 
onto the Riemann sphere and 
continuously deform two spheres by opening the branch cuts. 
We finally get a simple torus with $g=1$ with holes 
associated with the singularities.

\begin{figure}[htbp]
\begin{center}
\includegraphics[width=7cm]{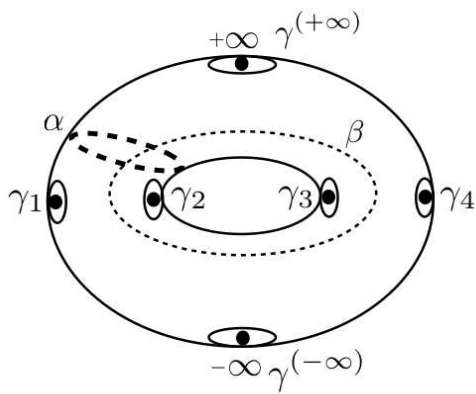}
\caption{
Homology basis of the fundamental group 
for the torus ${\cal T}\setminus \{ q_1,q_2,q_3,q_4, +\infty , -\infty \}$.}
\label{fundamentalloop}
\end{center}
\end{figure}

The homology basis of the fundamental group in this case 
is composed of the loops $\alpha,\beta$, which are homotopically independent loops on the torus,  
together with the loops encircling singularities. 
In addition to branch points at $q = q_i ~( 1\le i \le 4)$, 
there exist poles at $q = \pm \infty$, and we denote the loops 
associated with singularities by $\gamma_{i} ~ (1 \le i \le 4)$ and 
$\gamma^{(\pm \infty)}$, respectively  (see figure \ref{fundamentalloop}). 
Relations~(\ref{eq:homrelations}) allow to express, $\gamma_4$, say, as a product
of the other loops considered to be independent. 
As shown in the previous section, with fixed initial and final end points, 
the variety of distinct values of the action integral is 
given based on the formula (\ref{general_expression}). 


Recall the semiclassical formula (\ref{splitting}) for the tunnelling splitting 
requires the complex paths connecting the points symmetrically located in the 
$q$-plane 
then we may take~$\Gamma_0$ to connect $q_2$ and $q_3=-q_2$. 
Without loss of generality, we can obtain arbitrary symmetric paths 
from the path connecting the branch points $q_2$ and $q_3$ by shifting 
both initial and final points simultaneously keeping the symmetry condition. 
All the topologically distinct paths, taking into account the contribution 
from divergent singularities, are then written as 
\begin{eqnarray}
\Gamma = \Gamma_0+n_{\alpha}\alpha+n_{\beta}\beta
+\sum_{i=1}^{3}n_{\gamma_{i}}\gamma_{i}
+n^{(+\infty)} \gamma^{(+ \infty)} + n^{(-\infty)} \gamma^{(- \infty)} ,
\end{eqnarray}
where $n_{\alpha},n_{\beta},n_{\gamma_{i}}$ and $n^{(\pm\infty)}$ are winding numbers of each loop.

As is discussed below, 
it is important to specify the $\alpha$ and $\beta$ loops explicitly 
when one actually evaluates the action integrals, while 
we can freely move and deform the $\alpha$ and $\beta$ loops 
and the locations are not relevant 
within the argument of the fundamental group (see figure \ref{gfoldtorus}).

\begin{figure}[htbp]
\begin{center}
\includegraphics[width=8cm]{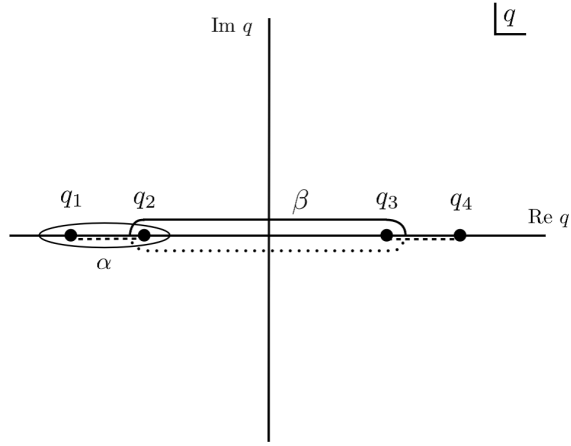}
\caption{$\alpha,\beta$ loops taken as integration contours on the $q$ plane.}
\label{abloop}
\end{center}
\end{figure}

For simplicity, 
we take two independent loops $\alpha$ and $\beta$ on the torus in such a way that 
each branch in the $\alpha$ loop runs along the real $q$-axis with encircling 
the two branch points $q_1$ and $q_2$, and in the same way 
the $\beta$ loop encircles the two branch points $q_2$ and $q_3$ 
(see figure \ref{abloop}). 
By taking the loops $\alpha$ and $\beta$ in this manner, 
the action integral for the $\alpha$ loop turns out to be real valued and that 
for the $\beta$ loop purely imaginary valued. 
As shown in Appendix B, the action integrals for $\gamma_{i}$ 
$(i=1,2,3)$ vanish. 
We then reach the expression for the total action integral after summing over all 
the contributions as
\begin{eqnarray}
S_{\Gamma}=S_{\Gamma_0}+n_{\alpha}S_{\alpha}+n_{\beta}S_{\beta}+
n^{(+\infty)} S^{(+ \infty)} + n^{(-\infty)} S^{(- \infty)} ,
\end{eqnarray}
where
\begin{eqnarray}
S_{\Gamma_0}&:=&\int_{q_2}^{q_3}pdq,\nonumber\\ 
S_{\alpha}&:=&\oint_{\alpha} p(q)dq=2\int_{q_1}^{q_2}pdq,\nonumber\\ 
S_{\beta}&:=&\oint_{\beta} p(q)dq=2\int_{q_2}^{q_3}pdq,\nonumber\\
S^{(\pm \infty)}&:=&\oint_{\gamma^{(\pm\infty)}} p(q)dq.
\end{eqnarray}

\begin{figure}[htbp]
\begin{center}
\includegraphics[width=10cm]{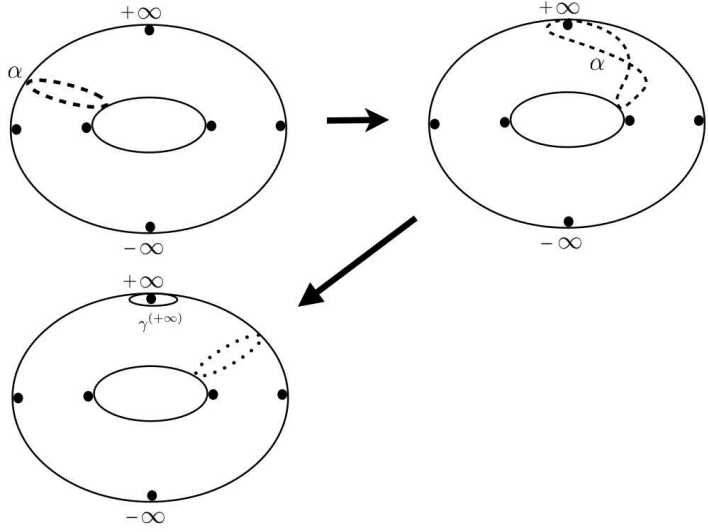}
\caption{Deformation of the $\alpha$ loop in the left well. 
It splits into 
a combination of the $\alpha$ loop in the right well and a loop around $+\infty$. }
\label{DeformDW}
\end{center}
\end{figure}

Now we show that $S_{\alpha}$ and $S^{(\pm \infty)}$ are not independent 
and actually related with each other. 
To see this, we rewrite as $S_{L} = S_{\alpha}$ for the left-side well, 
and introduce the action integral for the right-side well as
\begin{eqnarray}
S_{R}:=2\int_{q_3}^{q_4}pdq . 
\end{eqnarray}
As illustrated in figure \ref{DeformDW}, the integration contour specifying 
the action integral $S_{L}$ is continuously deformed and split into the ones 
associated with the action integrals $S^{(+\infty)}$ and $S_{R}$. 
This leads to the relation 
\begin{eqnarray}
S_{L} = S_{R} - S^{(+ \infty)}, 
\label{DWR}
\end{eqnarray}
where the minus sign in front of $S^{(+ \infty)}$ comes from the phase of $p$ (see Appendix C).
From the symmetry of the potential function, it is obvious that 
$S_{L}=S_{R}$ holds. This automatically gives $S^{(+ \infty)}=0$, 
which can also be confirmed by the direct calculation of 
the residue at $q=+\infty$ (also see Appendix C). 
From this observation, 
the candidates of action integrals finally take a simple form as
\begin{eqnarray}
S_{\Gamma}=S_{\Gamma_0}+n_{\alpha}S_{\alpha}+n_{\beta}S_{\beta}.
\label{actionexpressionDW}
\end{eqnarray}

Next we turn our attention to the most dominant complex path 
in the semiclassical formula (\ref{splitting}). 
Since classical action integrals under consideration are complex valued, 
the most dominant contribution is supposed to 
come from the complex classical 
orbit(s) with minimal imaginary action Im\,$S$. 
In the present situation, the $\alpha$ loop contribution is real valued, so the imaginary part of action integral is written as 
\begin{eqnarray}
{\rm Im}\, S_{\Gamma}={\rm Im}\, S_{\Gamma_0}+n_{\beta}{\rm Im}\, S_{\beta}.
\end{eqnarray}
This may take arbitrarily large negative values as $n_{\beta}$ is 
allowed to be any integer, positive or negative, meaning that 
imaginary action can become arbitrarily small. 
However, it is obvious that the orbits with negative 
imaginary action give rise to exponentially large contributions, 
which are not physically accepted, so should be dropped from 
the final contributions. 

Excluding unphysical contributions out of necessary ones could be done 
by handling the Stokes phenomenon properly. This would therefore 
be a matter of issues which should be closely discussed in order 
to make our theory self-consistent. 
However, as mentioned in section 2, 
we here treat the Stokes phenomenon only in a heuristic manner. 
The principle we adopt is based on the behavior 
of imaginary action as time proceeds. 
From the Hamiltonian equations of motion, $dq=pdt$ follows, which 
results in  $\int p dq = \int p^2 dt$. We then have 
\begin{eqnarray}
{\rm Im} \, S = \int -{\rm Im}\, p^2 {\rm Im}\, dt.
\end{eqnarray}
and in order to get~$\mathrm{Im} S>0$ we will choose a parametrisation such that~$\mathrm{Im}\, dt<0$.
In this choice, ${\rm Im}\,S$ becomes 
negatively large with increase in ${\rm Im}\, dt$ in a monotonic way.

If one applies this rule, which will also be used in the examples discussed below, 
$\Gamma_{0}$ is given as a trajectory passing through the potential barrier only once, that is a half cycle of the $\beta$ loop, and
the smallest imaginary action is just 
\begin{eqnarray}
{\rm Im}\, S_{\Gamma}=\frac{1}{2}\ S_{\beta}. 
\end{eqnarray}
This is nothing but the imaginary action for the so-called instanton path.
 From the expression (\ref{actionexpressionDW}), the corresponding real part turns out to be
\begin{eqnarray}
{\rm Re}\, S_{\Gamma}=n_{\alpha}S_{\alpha}.
\end{eqnarray}

Since branch points are turning points and $\alpha$, $\beta$ loops 
encircle the two branch points, we find that the Maslov index is equal to
 $\mu=2n_{\alpha}+1$.  Incorporating the semiclassical 
quantization condition $S_{\alpha}=(1/2+N)2\pi\hbar$, 
the formula (\ref{splitting}) can now be explicitly written  as 
\begin{eqnarray}
\frac{\hbar}{2T}e^{-S_{\beta}/2\hbar}\sum_{n_{\alpha}}(-1)^{2n_{\alpha}+2}(-1)^{n_{\alpha}}.
\end{eqnarray}
Here the sum over the winding number $n_{\alpha}$ is canceled except 
for the case $n_{\alpha}=0$. 
From these arguments we finally obtain the formula 
\begin{eqnarray}
\Delta E\sim \frac{\hbar}{2T}e^{-S_{\beta}/2\hbar}.
\end{eqnarray}
This is nothing but the well known formula 
 in the instanton theory, and also coincides 
with the result rederived in Ref. \cite{instanton}.

\section{Triple-well potential case}
\begin{figure}[htbp]
\begin{center}
\includegraphics[width=7cm]{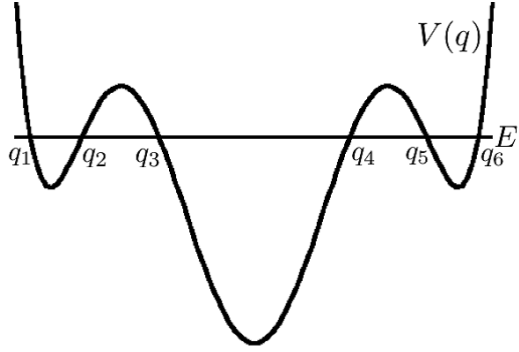}
\caption{The triple-well potential $V(q)$.}
\label{triplewellpot}
\end{center}
\end{figure}

As a next example, we consider a triple-well potential system:
\begin{eqnarray}
H(p,q)=\frac{p^2}{2}+V(q),
\label{TWpotential}
\end{eqnarray}
\begin{eqnarray}
V(q)=E+(q-q_1)(q-q_2)(q-q_3)(q-q_4)(q-q_5)(q-q_6),\nonumber
\end{eqnarray}
where the parameters $q_i ~ (1\le i \le 6)$ are all real 
and satisfy the conditions $q_1<q_2< \cdots <q_6$. 
We again assume the 
conditions $q_1=-q_6, q_2=-q_5, q_3=-q_4$ in order to develop the semiclassical analysis for 
the tunnelling splitting (see figure \ref{triplewellpot}). 

\begin{figure}[htbp]
\begin{center}
\includegraphics[width=12cm]{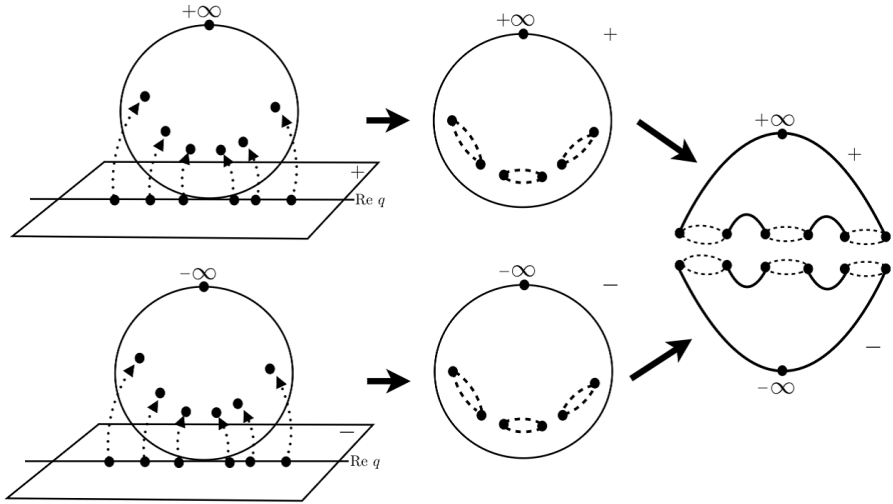}
\caption{
Deformation of Riemann spheres to a torus is shown in the triple-well potential case. The black dots and the dashed lines represent branch points and branch cuts, respectively. $\pm$ signs show the branches of $p(q)$.
}
\label{trance2}
\end{center}
\end{figure}

\begin{figure}[htbp]
\begin{center}
\includegraphics[width=8cm]{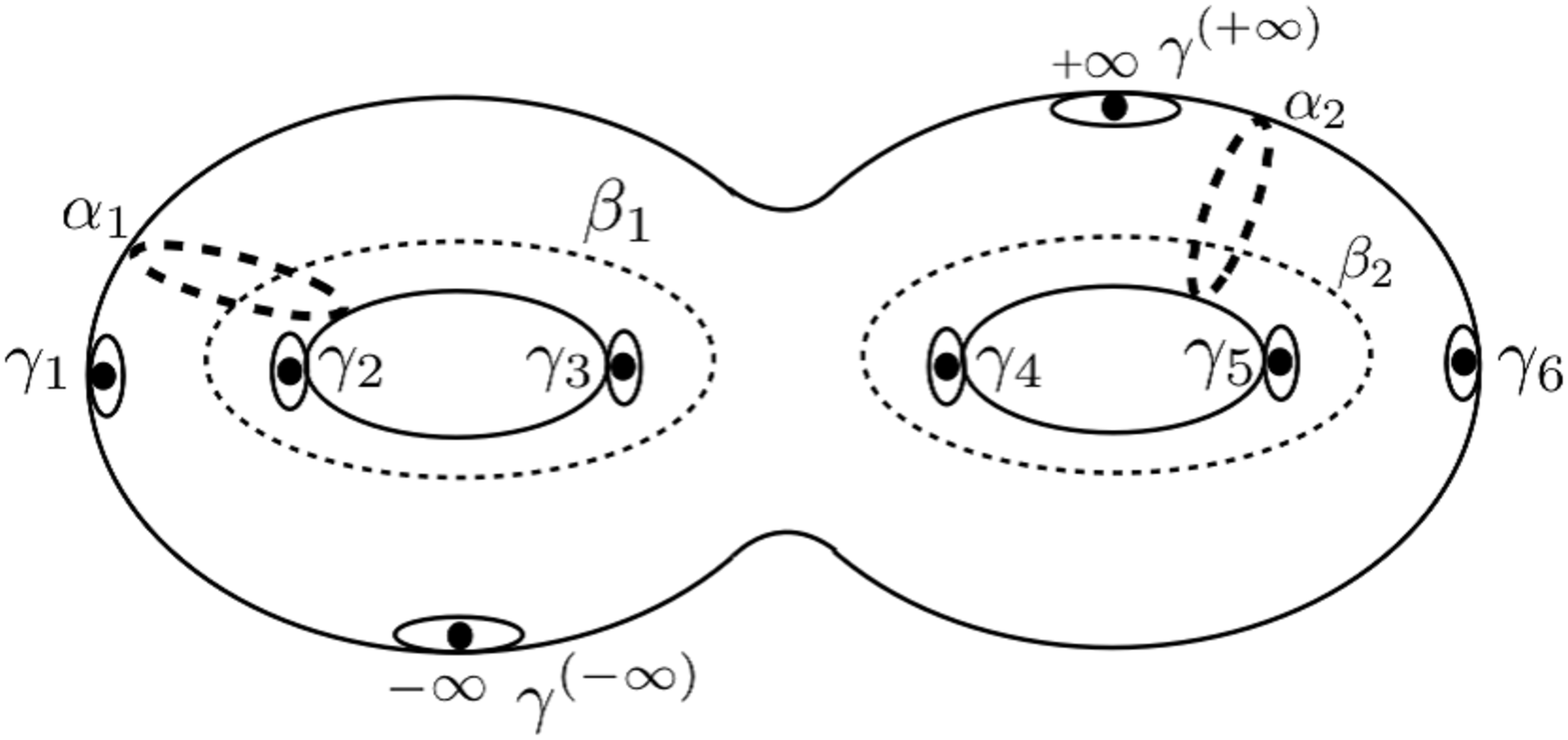}
\caption{Homology basis for the surface ${\mathcal T}\#{\mathcal T}\setminus \{q_{1},q_{2},q_{3},q_{4},q_{5},q_{6}, +\infty, -\infty \}$. 
}
\label{fundamentalloop2}
\end{center}
\end{figure}

In the same way as the double-well case, we obtain $p(q)$ as
\begin{eqnarray}
p(q)=\sqrt{-2(q-q_1)(q-q_2)(q-q_3)(q-q_4)(q-q_5)(q-q_6)}.\nonumber
\end{eqnarray}
The function $p(q)$ has now six branch points on the real axis. 
The associated Riemann surface of $p(q)$ is homeomorphic to a 2-fold torus
with small holes associated with branch points and poles (see figure \ref{trance2}).
The homology basis of the fundamental group is 
composed of the loops $\alpha_{i}$ and $\beta_{i}$ $(i=1,2)$ on the 2-fold torus and 
$\gamma_{i}$ $(1\le i \le 6)$ 
and $\gamma^{(\pm\infty)}$, each of which is a small loop 
encircling the corresponding singularity.
We illustrate in figure \ref{fundamentalloop2} 
the elements of the fundamental group
in this case.

To discuss the tunnelling splitting between 
the states localized at the left- and right-wells, 
let $\Gamma_0$ be a path connecting the branch points $q_2$ and $q_5=-q_2$. 
The integration contour is given by a combination of these loops as follows, keeping in mind that~$\gamma_6$
is a product of the other loops,
\begin{eqnarray}
\Gamma = 
\Gamma_0
&+& \sum_{i=1}^{2}n_{\alpha_{i}}\alpha_{i}+\sum_{i=1}^{2}n_{\beta_{i}}\beta_{i}\nonumber\\
&+& \sum_{i=1}^{5} n_{\gamma_{i}}\gamma_{i}
+n^{(+\infty)} \gamma^{(+ \infty)} + n^{(-\infty)} \gamma^{(- \infty)}, 
\end{eqnarray}
where $n_{\alpha},n_{\beta},n_{\gamma_{i}}$ and $n^{(\pm\infty)}$ are winding numbers of each loop. 
Again using the result shown in Appendix B, 
the action integrals for $\gamma_{i}$ $(i=1,2,\cdots, 5)$ all vanish, and  
we reach the expression for the total action integral 
contributions, 
\begin{eqnarray}
S_{\Gamma}&=&S_{\Gamma_0}+\sum_{i=1}^{2}n_{\alpha_{i}}S_{\alpha_{i}}+\sum_{i=1}^{2}n_{\beta_{i}}S_{\beta_{i}}
+
n^{(+\infty)} S^{(+ \infty)} + n^{(-\infty)} S^{(- \infty)},
\end{eqnarray}
where
\begin{eqnarray}
S_{\Gamma_0}&:=&\int_{q_2}^{q_5}pdq,\nonumber\\
S_{\alpha_{1}}&:=&\oint_{\alpha_{1}}pdq=2\int_{q_1}^{q_2}pdq,\nonumber\\
S_{\alpha_{2}}&:=&\oint_{\alpha_{2}}pdq=2\int_{q_5}^{q_6}pdq,\nonumber\\
S_{\beta_{1}}&:=&\oint_{\beta_{1}}pdq=2\int_{q_2}^{q_3}pdq,\nonumber\\
S_{\beta_{2}}&:=&\oint_{\beta_{2}}pdq=2\int_{q_4}^{q_5}pdq,\nonumber\\
S^{(\pm \infty)} &:=&\oint_{\gamma^{(\pm \infty)}} p(q)dq.
\end{eqnarray}

\begin{figure}[htbp]
\begin{center}
\includegraphics[width=12cm]{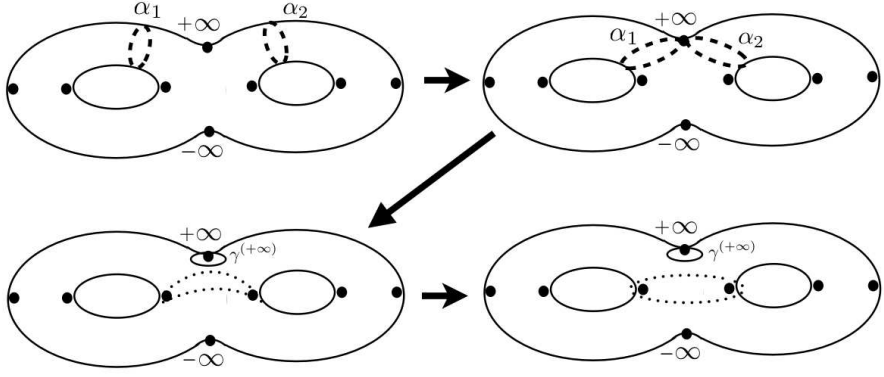}
\caption{Deformation of the $\alpha_1$ and $\alpha_2$ loops in the left- and right-wells. 
They split into 
a combination of the loop for the central well and a loop 
encircling $+\infty$. }
\label{Deform}
\end{center}
\end{figure}

As done in the double well potential case, we next show that 
these action integrals are not independent. 
As illustrated in figure \ref{Deform}, the integration contours specifying 
$S_{\alpha_{1}}$ and $S_{\alpha_{2}}$ are continuously 
deformed and split into the ones 
associated with the action integrals $S^{(+\infty)}$ and $S_{C}$. 
Here $S_{C}$ stands for the action integral for the central well, 
\begin{eqnarray}
S_{C} := 2\int_{q_3}^{q_4}pdq. 
\end{eqnarray}
Rewriting the notation as $S_L = S_{\alpha_1}$ and $S_R=S_{\alpha_2}$ 
to make clear that $S_{\alpha_{1}}$ and $S_{\alpha_{2}}$ are action 
integrals for the left- and right-side wells, 
we obtain the relation 
\begin{eqnarray}
S_C = S_{L}+S_{R} + S^{(+\infty)}.  
\label{actionrerationintriple}
\end{eqnarray}
This relation can also be confirmed in the direct calculation presented in 
Appendix C. A similar relation holds for $S^{(-\infty)}$ except that the sign in front of $S^{(-\infty)}$ is minus.

The symmetry of the potential function leads to 
the relations $S_{R}=S_{L}$, and $S_{\beta_{1}}=S_{\beta_{2}}$.  
As a result, all the possible classical action integrals are simply expressed as 
\begin{eqnarray}
S_{\Gamma}=S_{\Gamma_0}+n_{L}S_{L}+n_{C}S_{C}+(n_{\beta_{1}}+n_{\beta_{2}})S_{\beta_{1}}.
\end{eqnarray}
Note that the winding numbers are introduced as $n_{L}:=n_{\alpha_{1}}+n_{\alpha_{2}}-2n_{C}$ and $n_{C}:=n^{(+ \infty)}-n^{(- \infty)}$.

The principle to incorporate the Stokes phenomenon is the same as before. 
The imaginary part of complex paths is written as
\begin{eqnarray}
{\rm Im}\, S_{\Gamma}={\rm Im}\,S_{\Gamma_0}+(n_{\beta_{1}}+n_{\beta_{2}}){\rm Im} \, S_{\beta_{1}},
\end{eqnarray}
and we require that the imaginary component of time $t$ is decreasing. 
Under this condition, the complex path with the minimal imaginary action is 
given as the one with $n_{\beta_{1}}=n_{\beta_{2}}=0$. 
The corresponding orbit 
starts from the left-side well and crosses over two potential 
barriers and reaches the right-side well. The resulting imaginary action
is evaluated 
twice as much as the instanton action in each barrier:
\begin{eqnarray}
{\rm Im}\, S_{\Gamma}= {\rm Im} \, S_{\beta_{1}}.
\end{eqnarray}

Concerning the real part of the action integral, 
the path $\Gamma$ has to go half round the central well, 
so the real part of the action is given as 
\begin{eqnarray}
{\rm Re}\, S_{\Gamma}= n_{L}S_{L}+(n_{C}+\frac{1}{2})S_{C}, 
\end{eqnarray}
and the Maslov index is also evaluated similarly to give $\mu=2n_{L}+2n_{C}+3$.
We finally get the semiclassical expression for the tunnelling splitting:
\begin{eqnarray}
\Delta E_{n}\sim \frac{\hbar}{2T}e^{-S_{\beta_{1}}/\hbar}\sum_{n_{L},n_{C}}(-1)^{\mu+1}e^{\imat (n_{L}S_{L}+(n_{C}+\frac{1}{2})S_{C})/\hbar}.
\label{spliitingTW}
\end{eqnarray}
This almost coincides with the formula derived in Ref. \cite{instanton}, 
but the way of enumerating the paths differs from the one adopted there, 
so the form of the sum is slightly different. 
As also discussed in Ref. \cite{instanton}, the interference caused by the sum in the right-hand side 
gives rise to resonances, which generate a series of spikes in the $\Delta E$ vs $1/\hbar$-plot.
Such a phenomenon could be understood as the resonant tunnelling or the 
Fabry-P$\acute{\rm e}$rot effect in optics \cite{Bohm,CRC}.

\section{Simultaneous quantization}

As given in (\ref{DWR}) and (\ref{actionrerationintriple}) the action
integrals for the $\alpha$ loops in the fundamental group are related
through the action integral associated with the loop encircling
infinity.  These relations will invoke {\it simultaneous quantization}
of distinct wells. Simultaneous quantization in distinct wells has
been discussed in Ref. \cite{Leboeuf93}, and the result obtained above
is essentially the same as the one derived there in the double-well
potential case.

We first explain how simultaneous quantization is achieved 
in the double-well case. 
Suppose the action integral for the left-side well is quantized as 
$S_{L}=(1/2+m_L)2\pi\hbar$. From the relation (\ref{DWR}), 
the action for the right well is also quantized as 
$S_{R}=(1/2+m_R)2\pi\hbar$ if and only if the action integral around infinity 
satisfies the condition $S^{(\infty)}=2\pi\hbar m^{(\infty)}$, 
where $m_R, m_L$ and $m^{(\infty)}$ are integers. 

Concerning the triple-well system, 
the relation (\ref{actionrerationintriple}) among action integrals is not enough to give simultaneously quantization of $S_{L}$ and $S_{R}$ even if $S^{(\infty)}=2\pi\hbar m^{(\infty)}$ with integers $m^{(\infty)}$ is satisfied. 
However, if the potential is symmetric as assumed in section 5, 
$S_{L}$ and $S_{R}$ are quantized simultaneously since $S_{L} = S_{R}$ follows in such a case. 

Note that the relations (\ref{DWR}) and (\ref{actionrerationintriple}) hold 
among the $\alpha$ loops in the fundamental group. 
It would be natural to explore whether or not the relation involving $\beta$ loops exist, 
which might provide further constraints for action integrals. 
Integrals of algebraic functions along $\alpha$ or $\beta$ loops 
are called periods of Abelian integrals \cite{math3}. 
In a general argument of Abelian integrals, 
the period of Abelian integrals of the first kind has a relation as
\begin{eqnarray}
\left(\int_{\beta_{1}}\omega,\cdot\cdot\cdot,\int_{\beta_{g}}\omega \right)=\left(\int_{\alpha_{1}}\omega,\cdot\cdot\cdot,\int_{\alpha_{g}}\omega \right)T,
\label{periodrelation}
\end{eqnarray}
where $T$ is called the period matrix and $\omega$ is the Abelian differential of the first kind, 
respectively \cite{math2}. However, since the function $p(q)$ has poles in the Riemann surface, the relation among $\alpha$ or $\beta$ loops might not take a linear form as given in (\ref{periodrelation}). 
If the relation is linear, it would not provide an additional relation generating extra constraints concerning the quantization condition.

\section{Normal Form Hamiltonian}
In this section, we examine the case where 
the Hamiltonian is built from
more general normal forms and whose tunnelling splittings were semiclassically studied in Ref. \cite{JMS} 
in order
to investigate 
the validity of the so-called resonance-assisted tunnelling scenario (RAT) \cite{Brodier,ref1}. 
As shown below, equi-energy contours look like typical patterns 
observed in the Poincar$\acute{\rm e}$ section of phase space 
in two-dimensional nearly integrable systems.
In the following we consider a Hamiltonian of the form \cite{ref4} 
\begin{eqnarray}
H(p,q)=\sum_{k=1}^{n}a_{k}(p^{2}+q^{2})^{k}+\sum_{l,m}b_{l,m}q^{l}p^{m},
\label{NFgeneral}
\end{eqnarray}
where $a_{k}$ and $b_{l,m}$ are constants. 
Note that the $b$'s are not all independent and depend only on 2 real parameters.
The argument based on the fundamental group for algebraic functions holds, in
particular, the formula (\ref{general_expression}) for the path $\Gamma$. 

As shown in an example below, if the coefficient of the highest order of $p$ in the Hamiltonian does not 
depend on the variable $q$, the action integral along $\gamma_{i}$ loop turns out to
 be $0$ (see Appendix B). 

\begin{figure}[htbp]
\begin{center}
\includegraphics[width=7cm]{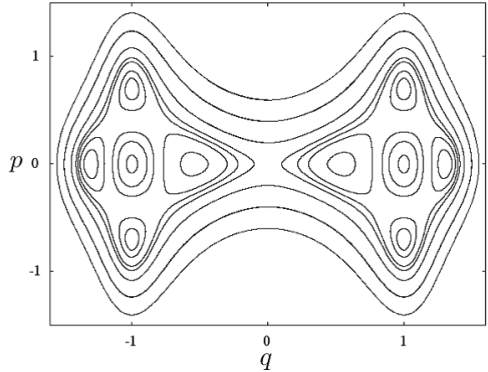}
\caption{Equi-energy contours for the Hamiltonian (\ref{NF}).}
\label{phaseNF}
\end{center}
\end{figure}

More specifically we will work with
\begin{eqnarray}
H(p,q)=\frac{1}{2}(p^{2}+x^{2})-\frac{1}{2}(p^{2}+x^{2})^2-2x^{2}p^{2},
\label{NF}
\end{eqnarray}
where $x:=1-q^{2}$. The symmetry condition~(\ref{eq:Hparity}) is maintained.
As seen in the phase space portrait drawn in figure \ref{phaseNF}, 
the system has two symmetric wells located at the positions $q=\pm 1$ respectively, 
and nonlinear resonance like equi-energy contours appear around each well. 

\begin{figure}[htbp]
\begin{center}
\includegraphics[width=9cm]{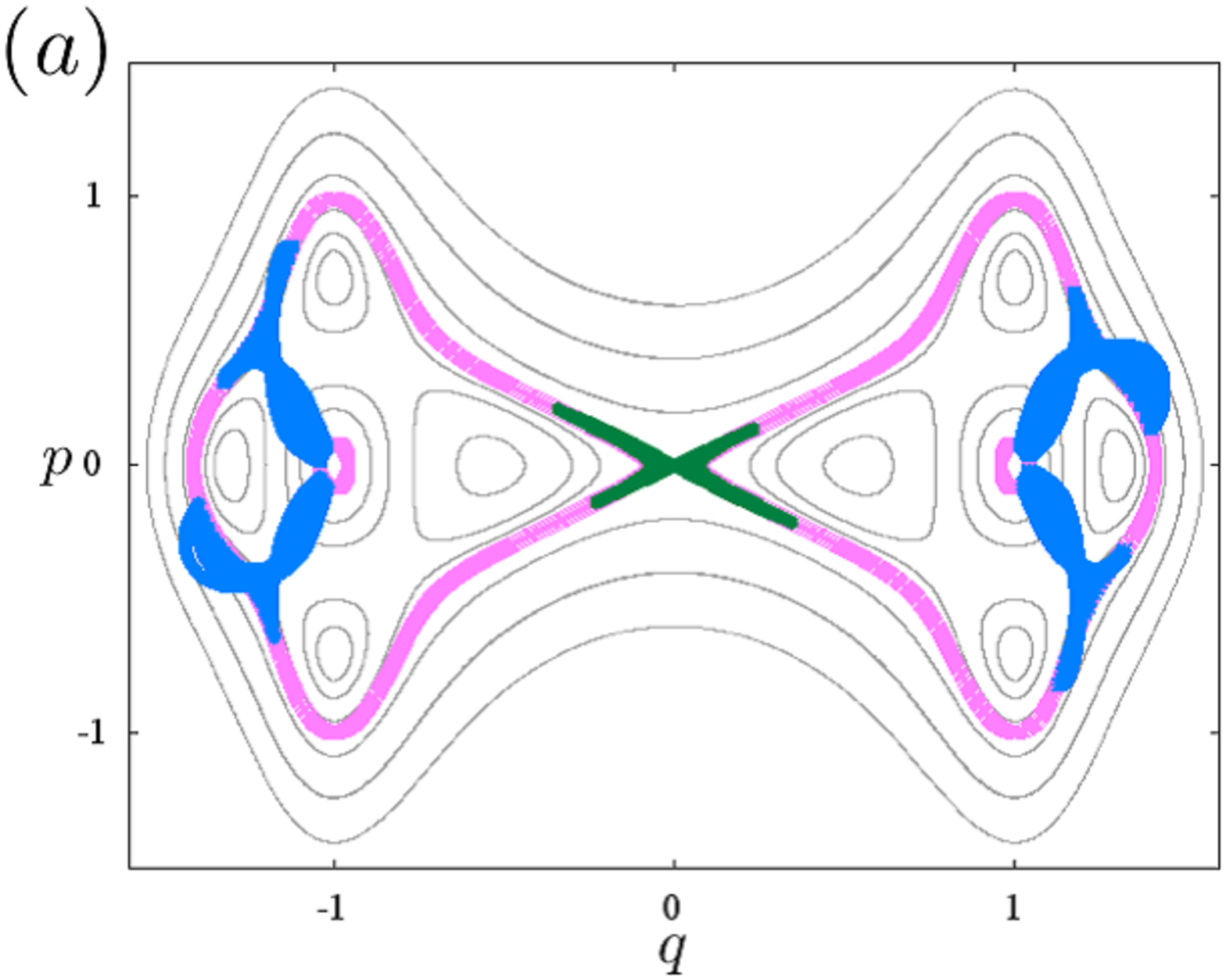}
\includegraphics[width=9cm]{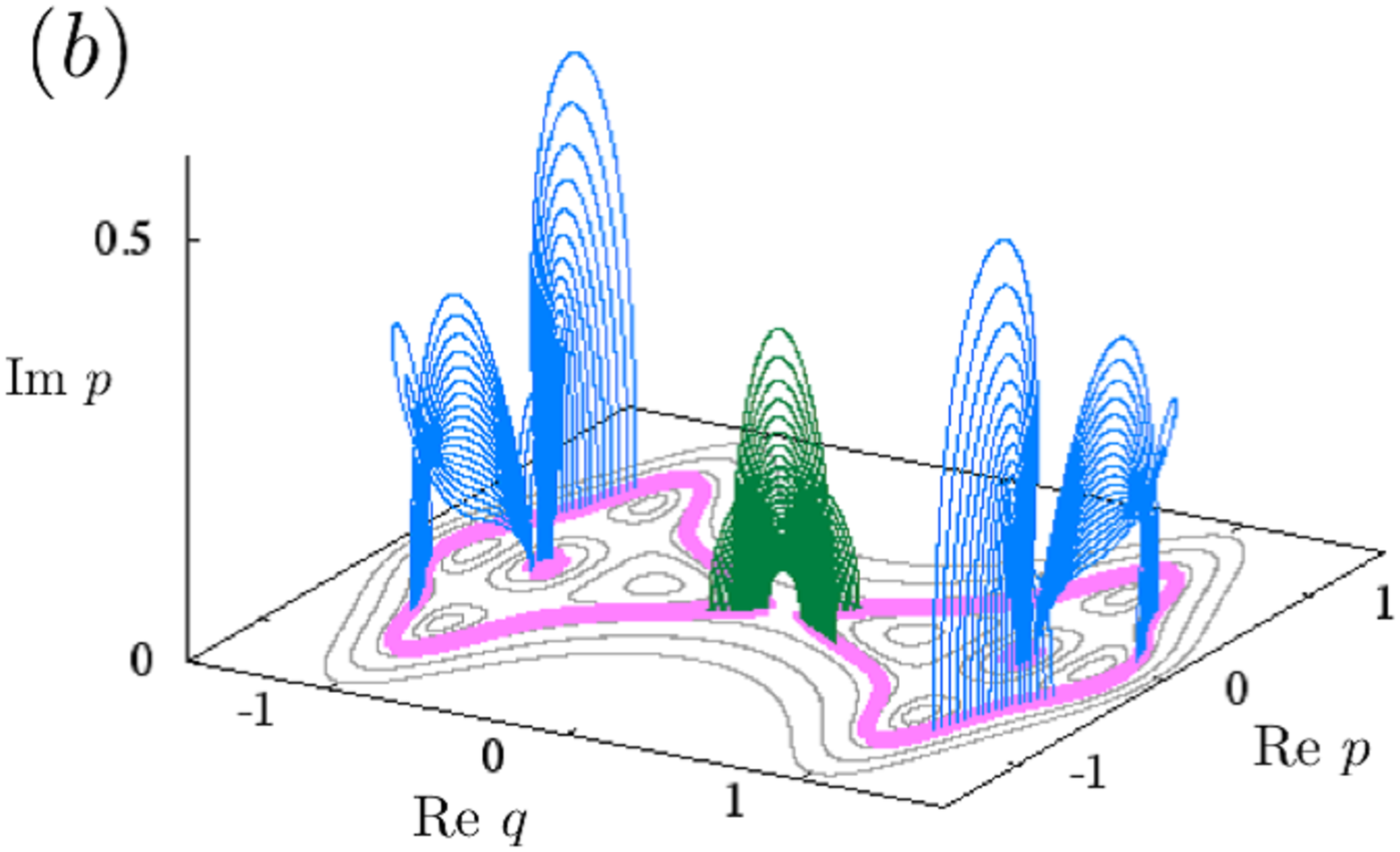}
\caption{(a) Equi-energy contours for the Hamiltonian (\ref{NF}).
The magenta curves show energy contours satisfying the 
condition $E\sim E_{n}^{\pm}$ where $E=6.19 \times 10^{-3}$.
The blue curves are projections onto the real plane  
of complex manifolds connecting 
inside and outside energy contours. 
The green one shows projection 
of complex manifolds connecting left and right outer energy contours. 
(b) The projection of each manifold onto $({\rm Re}\, q,{\rm Re}\, p,{\rm Im}\, p)$ space.} 
\label{manifolrNF}
\end{center}
\end{figure}

\begin{figure}[htbp]
\begin{center}
\includegraphics[width=7cm]{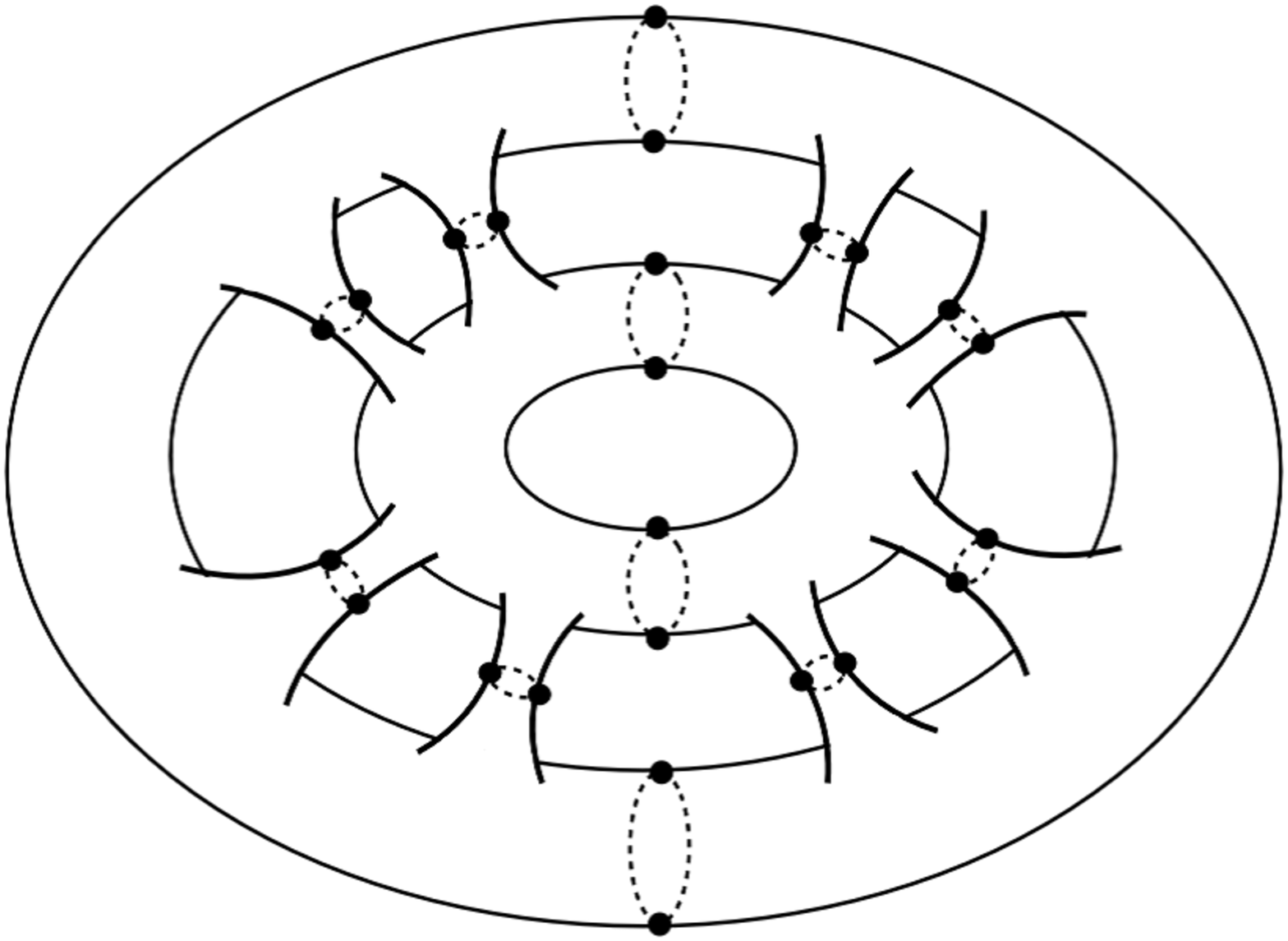}
\caption{The Riemann surface of $p(q)$  for the Hamiltonian (\ref{NF}).
The Riemann surface forms a $9$-fold torus. 
The black dots and dashed curves are branch points and branch 
cuts, respectively. }
\label{Phasespace}
\end{center}
\end{figure}

In order to perform semiclassical analysis for the tunnelling splitting, 
as was done in the previous examples, 
we first examine the Riemann surface and the associated fundamental group. 
From the Hamiltoninan (\ref{NF}), we easily find
\begin{eqnarray}\label{eq:pnormal}
p(q)=\sqrt{\pm\frac{\sqrt{-8E + 32 x^4 - 8 x^2+1}}{-2}-3x^2+\frac{1}{2}}. 
\end{eqnarray}
The branch points are obtained by solving simultaneous algebraic equations 
\begin{eqnarray}
\nonumber
&&\pm\frac{\sqrt{-8E + 32 x^4 - 8 x^2+1}}{-2}-3x^2+\frac{1}{2} =0, \\ \nonumber
&&-8E + 32 x^4 - 8 x^2+1 = 0, 
\end{eqnarray}
which provide 24 branch points in total. 
Each branch point is locally square-root type, thereby 
the corresponding Riemann surface $R$ has four leaves. 
Using the formula evaluating the genus, we find 
that the Riemann surface $R$  
is homeomorphic to $9$-fold torus with $28$ small holes associated 
with 4 poles and 24 branch points.
The Riemann surface is illustrated in figure \ref{Phasespace}. 
There are 9 $\alpha$- and $\beta$-loops together with 24 $\gamma$-loops
associated with the branch points and 4 $\gamma$-loops with poles,
each of which is attached in the corresponding leaf. 
From these observations, we have 45 independent action integrals in the semiclassical formula. 
However, as shown in Appendix B, the action integrals for $\gamma$-loops for 
branch points are all zero, and 
the residues at 
the poles 
vanish. 
This fact simplifies the expression of action integrals as 
\begin{eqnarray}
S_{\Gamma}=S_{\Gamma_0}+\sum_{i=1}^{9} n_{\alpha_{i}}S_{\alpha_{i}}+\sum_{i=1}^{9} n_{\beta_{i}}S_{\beta_{i}}.
\label{candidateinnormal}
\end{eqnarray}

The next step is to single out the most dominant path out of all the candidates given above. 
Since $dq=pdt$ does not hold any more, we cannot a priori compare the different~${\rm Im}\, S$ 
even with an increasing~${\rm Im}\, dt<0$ and the usual heuristic selection argument may fail, as will be shown below.

\section{Tunnelling splitting for the normal form Hamiltonian}

In the following, we discuss the tunnelling splitting $\Delta E_n$ for the normal 
form Hamiltonian (\ref{NF}) based on the 
semiclassical analysis. 
Note however that the semiclassical analysis performed here 
will not fully be based on the 
semiclassical formula (\ref{splitting}), and could be done 
only with a heuristic recipe. 
This is because, as shown below, that non-trivial situations
actually arise from the handling of the Stokes phenomenon, 
so the selection of the most dominant complex path would be 
highly non-trivial. 
We focus on the tunnelling splitting $\Delta E_n = E_n^- - E_n^+$ for~$E_n^+\simeq E_n^-\simeq E$ below the barrier. 
Figure \ref{INS} plots $\Delta E_{n}$ as a 
function of $1/\hbar$ for~$E=6.19 \times 10^{-3}$.

In classical phase space, there appear 
congruent energy contour pairs in both sides of equi-energy contours, reflecting 
the symmetry with respect to the $q$-direction. 
For the energy satisfying $E\sim E^{\pm}_{n}$, there appear two closed energy contours 
in each side, 
which are shown in magenta curves in figure \ref{manifolrNF}. Obviously, due to the symmetry, there are only two characteristic real periods $T_{{\rm out}}$ and~$T_{{\rm in}}$ and two actions~$S_{{\rm out}}$ and~$S_{{\rm in}}$
associated with the outer and inner orbits respectively. The latter are connected via complex manifolds, 
which are shown in blue curves in figure \ref{manifolrNF}, and outer periodic orbits in both sides are 
also connected via complex manifolds, drawn in green curves. 
Complex manifolds are obtained by integrating Hamiltonian equations of motion 
in the purely imaginary direction starting from each point of periodic orbits.

Following the argument developed in Ref. \cite{instanton}, 
we consider the time path in the complex plane for 
the orbit contributing to the semiclassical formula (\ref{splitting}). 
The total elapsed time $T$ is written as 
\begin{eqnarray}
T\sim R(T)+\imat T_{{\rm in}-{\rm out}} 
+\imat T_{{\rm out}-{\rm out}}
+\imat T_{{\rm in}-{\rm out}}+L(T),
\end{eqnarray}
where $\imat T_{{\rm in}-{\rm out}}$ is the time interval during which the complex orbit 
runs from the inner energy to the outer energy curve within the same well, 
shown in blue curves in figure \ref{manifolrNF}. 
Similarly, $\imat T_{{\rm out}-{\rm out}}$ is the purely imaginary interval between the outer energy curve 
in the left side to another outer curve in the right side, shown in green curves in figure \ref{manifolrNF}. 
$L(T)$ and $R(T)$ 
are sums of time intervals spent by the orbit moving in the inner and 
outer real energy curves, i.e.,
\begin{eqnarray}
L(T)&=&n_{{\rm in}}T_{{\rm in}}+n_{{\rm out}}T_{{\rm out}},
\label{TL}\\
R(T)&=&n'_{{\rm in}}T_{{\rm in}}+n'_{{\rm out}}T_{{\rm out}}, 
\label{TR}
\end{eqnarray}
where the winding numbers $n_{{\rm in}},n_{{\rm out}},n'_{{\rm in}}$ and $n'_{{\rm out}}$ are taken to be positive integers. 
A comment concerning $L(T)$ and $R(T)$ is in order. 
In Ref. \cite{instanton}, a fractional time interval $\tau$, or a residual time, was introduced 
for the time interval along the real direction as ${\rm Re}\, T=L(T)+R(T)-\tau$
in order to adjust the time interval in such a way that initial and final points are located at desired positions.
However this residual time $\tau$ does not play any roles after taking 
the limit ${\rm Re}\, T\rightarrow \infty$ \cite{instanton}. 
The corresponding action integral is then written as 
\begin{eqnarray}
S&\sim&n_{{\rm in}}S_{{\rm in}}+\imat S_{{\rm in}-{\rm out}}/2+n_{{\rm out}}S_{{\rm out}}\nonumber\\
&+&\imat S_{{\rm out}-{\rm out}}/2\nonumber\\
&+&n'_{{\rm in}}S_{{\rm in}}+\imat S_{{\rm in}-{\rm out}}/2+n'_{{\rm out}}S_{{\rm out}},
\end{eqnarray}

Here we focus only on the trajectories running on the complex manifolds connecting 
the real energy curves only once, as illustrated in figure \ref{TRA}. 
Hence,
under the restrictions given in (\ref{TL}) and (\ref{TR}), 
the sum of contributions of such trajectories takes the form as 
\begin{eqnarray}
&&\sum_{n_{{\rm in}}}\sum_{n'_{{\rm in}}}(-1)^{\mu+1}
4(n_{{\rm in}}+1)4(n'_{{\rm in}}+1) e^{\imat (n_{{\rm in}}S_{{\rm in}}+\imat S_{{\rm in}-{\rm out}}/2+n_{{\rm out}}S_{{\rm out}})/\hbar}\nonumber\\
&&\times e^{(-S_{{\rm out}-{\rm out}}/2)/\hbar}
e^{\imat (n'_{{\rm in}}S_{{\rm in}}+\imat S_{{\rm in}-{\rm out}}/2+n'_{{\rm out}}S_{{\rm out}})/\hbar}.
\end{eqnarray}
Here the Maslov index is evaluated as $\mu=n_{{\rm in}}+n_{{\rm out}}+n'_{{\rm in}}+n'_{{\rm out}}+7$. 
We may take the sums for $n_{{\rm in}}$ and $n'_{{\rm in}}$ separably, and 
each sum is the same as the one in the triple-well case in Ref. \cite{instanton}.
These lead us to the semiclassical expression for the tunnelling splitting 
\begin{eqnarray}
\Delta E_n \sim 
\frac{2\hbar}{T_{{\rm in}}}\left(\frac{e^{-S_{{\rm in}-{\rm out}}/(2\hbar)}}{\sin(((T_{{\rm out}}/T_{{\rm in}})S_{{\rm in}}-S_{{\rm out}})/(2\hbar))}\right)^2 e^{-S_{{\rm out}-{\rm out}}/2\hbar}.
\label{expandJM}
\end{eqnarray}

\begin{figure}[htbp]
\begin{center}
\includegraphics[width=8cm]{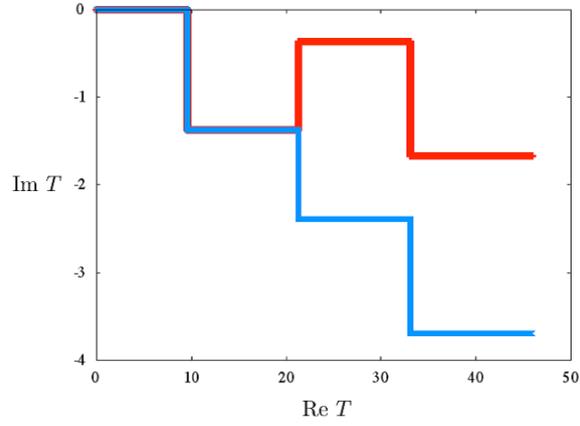}
\caption{Complex time paths which are taken to test the semiclassical 
formula (see text). 
The imaginary time monotonically decreases in the blue path case 
while monotonicity condition is not satisfied in the red path case. 
}
\label{timepath}
\end{center}
\end{figure}

\begin{figure}[htbp]
\begin{center}
\includegraphics[width=8cm]{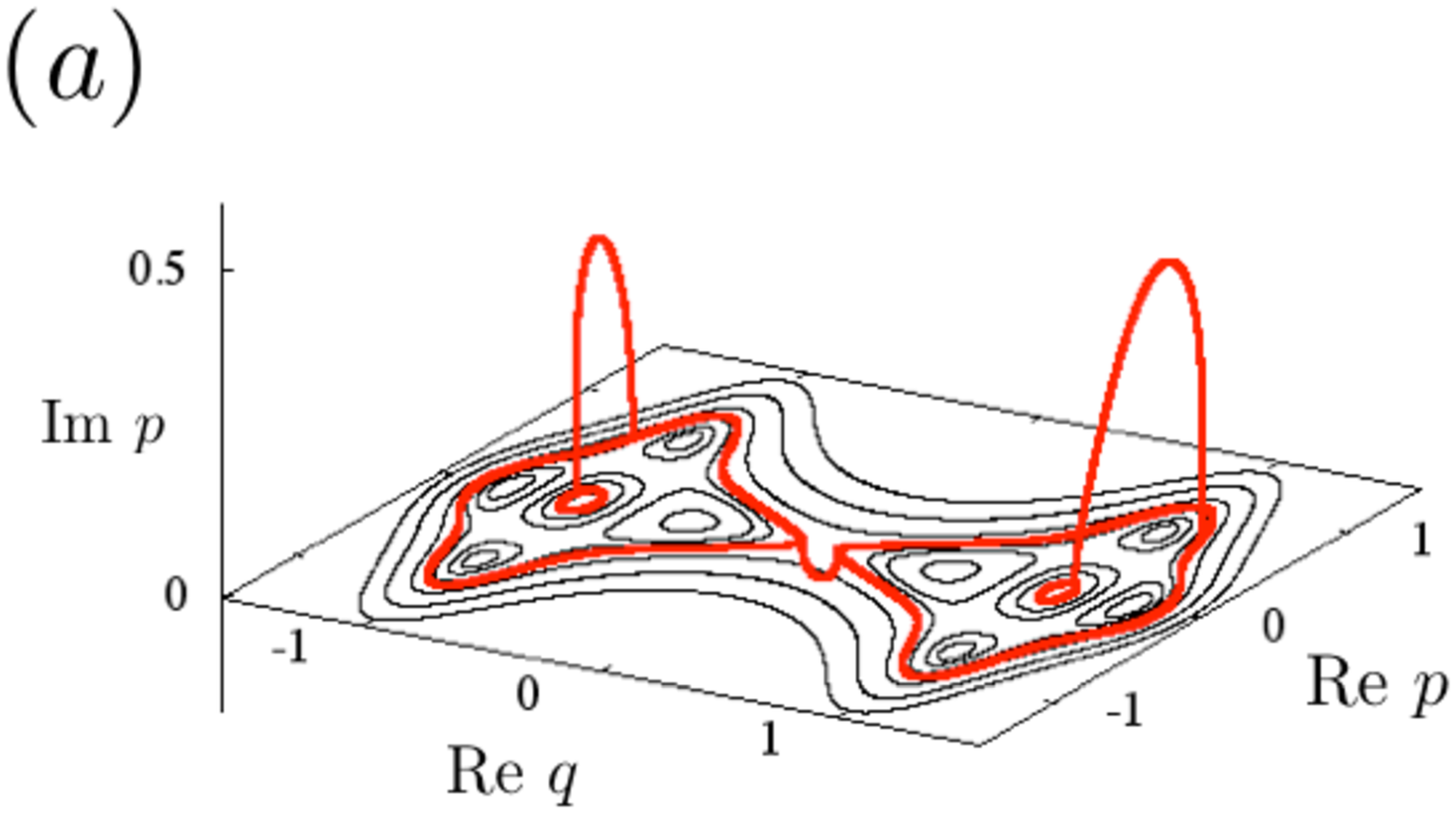}
\includegraphics[width=8cm]{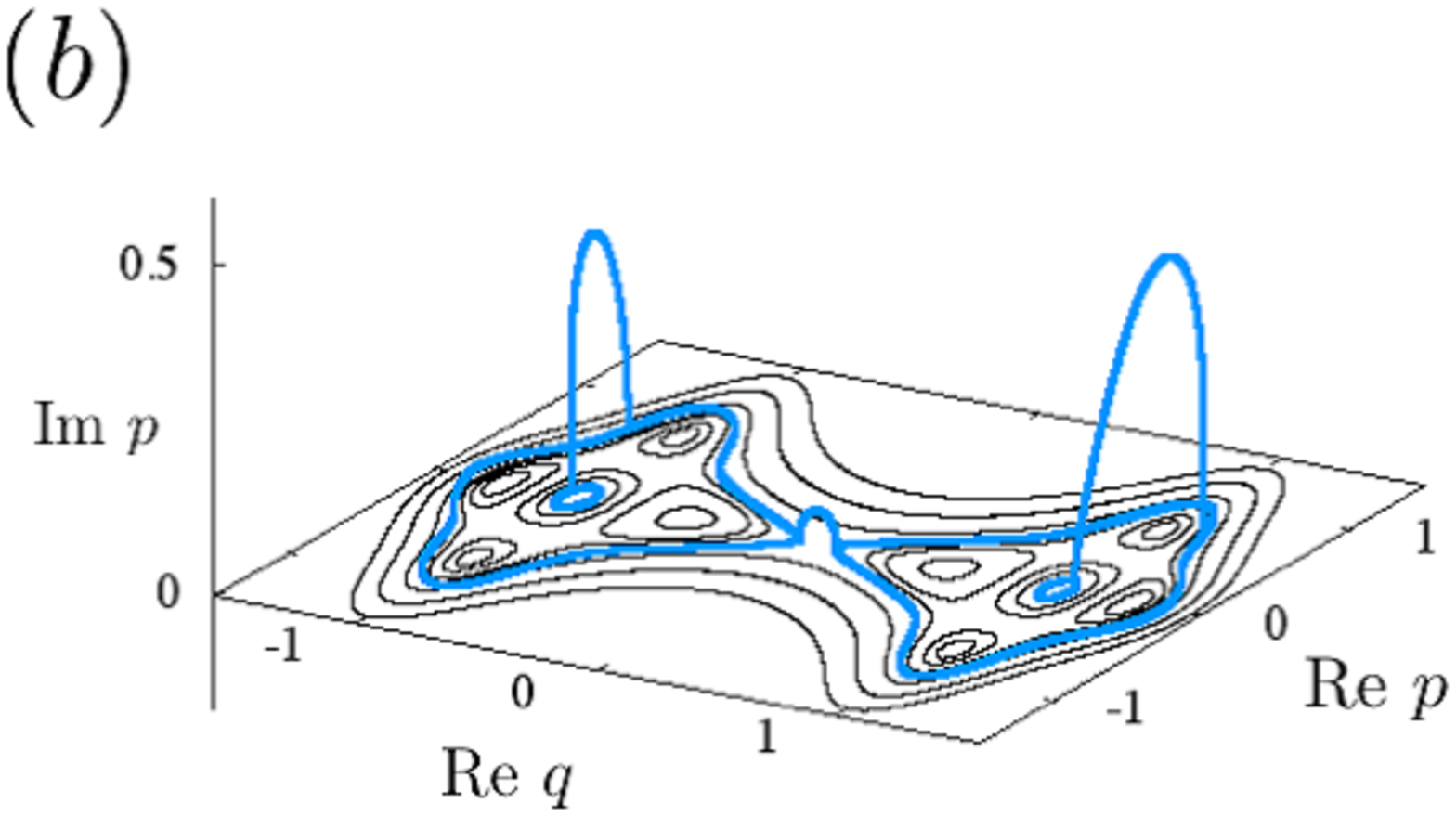}
\caption{The projection of complex paths onto $({\rm Re}\ q,{\rm Re}\ p,{\rm Im}\ p)$ space 
in the case (a) where the corresponding time path is taken as the red line in figure \ref{timepath} 
and (b) where the blue path is taken, respectively. 
The energy for red and blue colored curves is given as $E=6.19 \times 10^{-3}$.}
\label{TRA}
\end{center}
\end{figure}

\begin{figure}[htbp]
\begin{center}
\includegraphics[width=10cm]{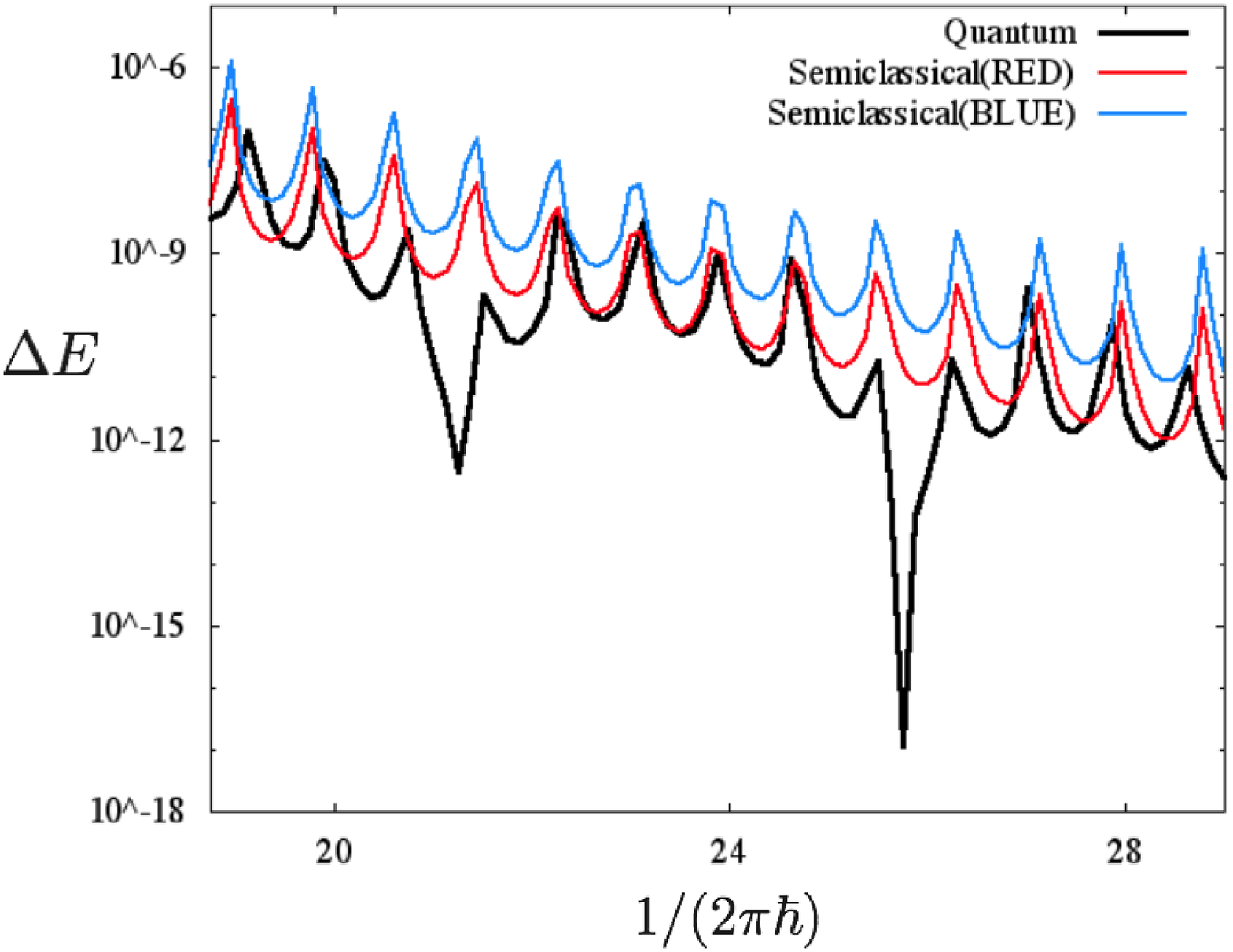}
\caption{
The tunnelling splitting $\Delta E_n = E_n^- - E_n^+$ as a function of $1/\hbar$. 
Here $E_n^+\simeq E_n^-\simeq E =6.19 \times 10^{-3}$. 
The black curve shows the numerical result obtained by direct calculation. 
The blue and red ones are obtained by applying 
the semiclassical formula (\ref{expandJM}), and 
the corresponding time paths are respectively shown in figure \ref{timepath}. 
}
\label{INS}
\end{center}
\end{figure}

Using this formula, we now demonstrate that a proper treatment of 
the Stokes phenomenon is crucial to discuss the tunnelling 
splitting of the normal form Hamiltonian within the semiclassical framework. 
In figure \ref{INS}, we compare the splitting calculated using direct diagonalization 
with the ones obtained using the semiclassical formula (\ref{expandJM}). 
In the semiclassical calculation, we show the splittings evaluated  
using the complex path, which are drawn as red and blue zig-zag lines in the complex time plane (see figure \ref{timepath}). 
Note that both paths connect the left- and right wells and satisfy 
the boundary conditions necessary for the semiclassical formula. 

As noticed from figure \ref{timepath}, 
the time path shown in blue satisfies the condition that Im\,$T$ monotonically 
decreases whereas the path in red breaks the monotonicity. 
According to the criterion adopted in sections 4 and 5, 
the red-colored path should be dropped from the final 
contribution because the path contains an interval in which 
Im\,$T$ increases and expected to provide an exponentially exploding 
contribution which should be excluded from the final sum. 
However, as seen in figure \ref{INS}, 
the curve based on the red-path contribution gives a larger slope 
as compared to the blue one, and shows better fitting to the exact plot.
This result provides evidence implying that a naive criterion to treat the 
Stokes phenomenon does not work in the case studied here. 
The result also strongly suggests that exponentially decreasing solutions 
do not necessarily remain as contributions. 
This is counterintuitive in the conventional semiclassical argument as well.


\section{Summary and Discussion}

In this paper we have investigated the topology of complex paths 
in one-dimensional systems 
to enumerate possible complex paths which contribute to the semiclassical 
sum formula for tunnelling splittings.
Here Hamiltonian functions were assumed to be written as polynomials of 
the variables $p$ and $q$, thereby we could make use of knowledge on 
algebraic functions, especially the fundamental group for the Riemann surface. 
Since the action integral is the most important ingredient in the semiclassical formula, 
we examined the Riemann surface of the function 
$p(q)$ 
closely and showed that it has a finite number of leaves and homeomorphic to a 
multi-handled compact surface. 
The number of loops of the homology basis for the associated fundamental 
group turns out to be finite, reflecting the fact that the function $p(q)$ is algebraic. 

To enumerate independent action integrals, 
it would be natural to consider independent elements in the fundamental group 
of the function $p(q)$. However this is not enough for our semiclassical analysis 
because the action 
integral is defined by the integration of $p(q)$ along an integration contour, 
so one has to take into account not only branch points generating the multivaluedness 
of the function $p(q)$, but also other singularities of $p(q)$ with non-zero 
residues. 
Such singularities indeed appear in the Riemann surface 
as divergent points of $p(q)$. 

As model systems, we here studied the double- and triple-well potential systems, 
together with the normal form Hamiltonians as well. 
For the former two cases, we have obtained the complete list 
of the possible complex paths based on the idea employing the fundamental group. 
As a bi-product out of such a systematic treatment,
we derived action relations involving the residue contribution from divergent points of $p(q)$. 
Note that the relation for the double-well case 
has already been obtained in \cite{Leboeuf93}, but its origin could  
more simply be understood through the fundamental group argument. 
In the case of the double-well potential system for instance, 
we usually consider the quantization condition for each well independently 
since the equi-energy surfaces in left- and right-side wells are classically disjoined.
However our analysis exploring the topology of the whole complex equi-energy 
surfaces has unveiled that 
quantization conditions in left and right wells 
are linked through 
the action integral associated with infinity of the Riemann surface. 
Similar action relations were similarly derived in the triple-well potential system, 
and they lead to simultaneous quantization of left- and right-wells
if the potential is symmetric. 

In performing the semiclassical analysis,  it is not sufficient to enumerate the complex 
paths satisfying the boundary conditions required in the semiclassical formula. 
Since the semiclassical formula is obtained by applying 
the saddle point method, one needs to 
handle the Stokes phenomenon in an appropriate manner. 
In the semiclassical arguments for tunnelling splittings so far, 
this issue has not been discussed seriously 
even in one-dimensional situations. 
The most typical approach would be just to remove exponentially exploding 
solutions, which is based only on a rather naive speculation in 
analogy with a treatment of the Airy function. 
The well-known instanton theory and its variants applied to more general 
situations have adopted essentially the same strategy.
However, as shown in the present paper, the possible classical actions are expressed as 
a linear combination of elements of the fundamental group together with 
contributions from divergent singularities. 
This brings infinitely many possible candidates, and 
infinitely many exploding solutions are necessarily contained among them.  
As a result, it becomes a crucial step to deal with the Stokes phenomenon 
properly. 
This is entirely beyond the scope of this paper, and here we only tested 
the most conventional prescription. 
For the double- and triple-well potential systems, 
we extracted the complex paths remaining as semiclassical contributions 
in such a way that the imaginary direction of 
the corresponding time path should be negative, which guarantees 
the monotonicity of imaginary action of complex paths. 
We confirmed that 
the results were both consistent with known results.

In the normal form Hamiltonian case, 
we could also find all the possible complex paths based on the 
fundamental group because the Hamiltonian is also given as a polynomial function. 
However, a naive treatment of the Stokes phenomenon was shown to break down. 
In particular, we demonstrated that there is a situation where 
even exponentially decaying contributions should be dropped, which 
is one piece of evidence suggesting that 
the Stokes phenomenon for the normal form Hamiltonian systems 
must be highly non-trivial \cite{SI16}. 

Our motivation for studying the normal form Hamiltonian was to 
promote our understanding of the so-called resonance-assisted tunnelling 
as was done in Ref. \cite{JMS}. 
As stressed in this paper, equi-energy contours of the normal form 
Hamiltonian apparently look like patterns typically appearing 
in Poincar$\acute{\rm e}$ sections 
of two-dimensional nearly integrable system, 
but nonlinear resonance like structures in one-dimensional systems 
are not caused by nonlinear resonances. 
It would therefore be unreasonable to explain the mechanism of tunnelling 
occurring in two-dimensional nonintegrable systems based on
one-dimensional systems even though apparent similarity exists in 
their phase space patterns. 

Even if one concedes that the normal form Hamiltonian could somehow serve 
as an analogous model to the system with nonlinear resonances, 
the analysis based on the fundamental group tells us that 
what is relevant is the topology of the Riemann surface, which is 
entirely controlled by the branch points of the function $p(q)$. 
This implies that instanton in the conventional sense might play only a relative role. 
The understanding of instanton has been updated 
from the perspective of the relevance of the Riemann sheet structure, 
which is based on a similar spirit as our present arguments \cite{Dulden,Dulden2}.
 
One important message out of this paper would be that one does not need to consider 
the time path any more and has only to focus on the function $p(q)$. 
Instanton has a long history and the idea using the complex time plane 
has been and still might be predominant, but we believe that this would not be a right strategy as
discussed in the present paper and Refs. \cite{Dulden,Dulden2} as well. 
What we need is information on the function $p(q)$, not 
the complex structure of functions $q(t)$ and $p(t)$, 
so analyzing the fundamental group for the Riemann surface of $p(q)$
would become unavoidable. 
\\
\\
\\
\\
\\
{\Large\bf Acknowledgments}\\
The authors are very grateful to Masanori Kobayashi for many 
valuable comments on the fundamental group for 
algebraic functions. This work has been supported by JSPS KAKENHI Grant Numbers 25400405 and 15H03701.

\appendix
\section{Semiclassical formula of tunnel splittings}

In this appendix, we briefly sketch the derivation of the formula (\ref{splitting}), following 
Ref. \cite{instanton}. 
Let $|\phi^{\pm}_{n}\rangle$ be symmetric and asymmetric quasi-degenerated states 
for a Hamiltonian commuting with the parity operator $\hat S$ such that $\hat S^2 = 1$. The eigenstates of~$\hat{H}$ can be classified according to their parity, 
\begin{eqnarray}
\hat S|\phi^{\pm}_{n}\rangle\DEF\pm|\phi^{\pm}_{n}\rangle.
\end{eqnarray}
The spectral decomposition of the evolution operator after a time~$T$ writes
\begin{eqnarray}
 \hat U(T) =\sum_{n} (e^{\frac{1}{\imat\hbar}E^{+}_{n} T}|\phi^{+}_{n}\rangle \langle \phi^{+}_{n}|+e^{\frac{1}{\imat\hbar}E^{-}_{n} T}|\phi^{-}_{n}\rangle \langle \phi^{-}_{n}|).
\end{eqnarray}
To discuss the tunnelling splitting between the states $|\phi^{+}_{n}\rangle$ and $|\phi^{-}_{n}\rangle$, we further define the projection operator 
\begin{eqnarray}
 \hat \Pi_{n} \DEF|\phi^{+}_{n}\rangle \langle\phi^{+}_{n}|+|\phi^{-}_{n}\rangle \langle\phi^{-}_{n}|.
\end{eqnarray}
and we have 
\begin{eqnarray}
{\rm Tr}( \hat \Pi_{n} \hat U) = \sum_{m} \langle \phi^{\pm}_{m}|\hat \Pi_{n} \hat U|\phi^{\pm}_{m}\rangle
=e^{-\frac{\imat}{\hbar}E^{+}_{n} T}+e^{-\frac{\imat}{\hbar}E^{-}_{n} T},
\end{eqnarray}
\begin{eqnarray}
{\rm Tr}(\hat S\hat \Pi_{n}\hat U) =e^{-\frac{\imat}{\hbar}E^{+}_{n} T}-e^{-\frac{\imat}{\hbar}E^{-}_{n} T}.
\end{eqnarray}
We then obtain 
\begin{eqnarray}
\frac {{\rm Tr}(\hat S \hat \Pi_{n}\hat U)}{{\rm Tr}(\hat \Pi_{n} \hat U)} 
=\imat\tan(\frac{\Delta E_{n}}{2\hbar} T). 
\end{eqnarray}
where $\Delta E_{n}=E^{-}_{n}-E^{+}_{n}$. 
If the condition 
\begin{eqnarray}
\frac{|T|\Delta E_{n}}{2\hbar} \ll 1
\end{eqnarray}
is satisfied, the tunnelling splitting $\Delta E_{n}$ can be explicitly written as 
\begin{eqnarray}
\Delta E_{n} \sim \frac{2\hbar}{\imat T}\frac {{\rm Tr}(\hat S \hat \Pi_{n}\hat U)}{{\rm Tr}( \hat \Pi_{n}\hat U)}.
\label{deltaE}
\end{eqnarray}

We now rewrite the right-hand side of (\ref{deltaE}) in the path integral form. 
Introducing the quasi-mode $|\Phi_{n}\rangle\DEFt(|\phi_{n}^{+}\rangle+|\phi_{n}^{-}\rangle)/\sqrt{2}$, 
the projection operator is expressed as 
\begin{eqnarray}
|\phi_{n}^{+}\rangle \langle \phi_{n}^{+}|+|\phi_{n}^{-}\rangle \langle \phi_{n}^{-}|=|\Phi_{n}\rangle \langle \Phi_{n}|+\hat S|\Phi_{n}\rangle \langle \Phi_{n}|\hat S.
\end{eqnarray}
Let $\Phi_{n}^{sc}(q)$ be WKB approximation of $\Phi_{n}(q)$ \cite{Keller,percival}, 
which is localized on the energy curve satisfying $E\sim E^{\pm}$, then 
the numerator and denominator of the formula (\ref{deltaE}) are 
semiclassically evaluated as 
\begin{eqnarray}
2\int dqdq' \Phi_{n}^{sc}(q)(\Phi_{n}^{sc}(q'))^{*}G(\eta q',q;T),
\label{WKBGVV}
\end{eqnarray}
where $G(\eta q',q;T)$ represents the Van Vleck-Gutzwiller propagator 
\begin{eqnarray}
G(\eta q',q;T)=\sum_{\gamma} (-1)^{k_{\gamma}}\sqrt{\det\left(\frac{\imat}{2\pi\hbar}\frac{\partial^{2}S_{\gamma}}{\partial q_{f}\partial q_{i}}\right)}e^{\frac{\imat}{\hbar}S_{\gamma}(\eta q',q;T)}. 
\label{VV}
\end{eqnarray}
Here $\eta=-1$ for the numerator and $\eta=+1$ for the denominator of the formula (\ref{deltaE}), 
respectively. 
The index $k_{\gamma}$ denotes the number of conjugation points along the trajectory $\gamma$.

We further evaluate the integral (\ref{WKBGVV}) again using the saddle point approximation, 
which requires the condition 
\begin{eqnarray}
\lim_{q_{f}\rightarrow \eta q_{i}}{\frac{\delta S_{\gamma}}{\delta q_{i}}}=\frac{\partial S_{\gamma}}{\partial q_{i}}+\frac{\partial q_{f}}{\partial q_{i}}\frac{\partial S_{\gamma}}{\partial q_{f}}=0.
\end{eqnarray}
Then the generating relations 
\begin{eqnarray}
\frac{\partial S_{\gamma}}{\partial q_{i}}=-p_{i},\ \ \ \ \ \ \frac{\partial S_{\gamma}}{\partial q_{f}}=\eta p_{f},
\end{eqnarray}
leads to the condition 
\begin{eqnarray}
p_{f}=\eta p_{i}
\end{eqnarray}
for each $\eta$. 
By taking the trace of integral (\ref{WKBGVV}), 
the classical paths contributing to the final semiclassical sum  
should altogether satisfy the conditions $E\sim E_{n}^{\pm}$, $q_{f}=\eta q_{i}$
and $p_{f}=\eta p_{i}$.
In section 2, $q_{f}$ and $q_i$ are expressed as $q(T)$ and $q(0)$, respectively (same as for $p$).
After calculating the prefactor in evaluating the integral (\ref{WKBGVV}) (see details in Ref. \cite{instanton}), 
we finally reach the formula (\ref{splitting}).

\section{Integral along $\gamma$ loops}

In this appendix we calculate the integral whose integration contour 
encircles a single branch point $q_{i}$ of the function $p(q)$. 
In the text, such a loop is called the $\gamma_i$ loop. 

Branch points of the algebraic function are algebraic singularities around which~$p$ has the Puiseux expansion 
in the following form 
\begin{equation}
p(q) = \sum_{n=s}^{\infty}c_{n}(q-q_i)^{\frac{n}{w}}, \ \ \ \ \ \ (s>-\infty)
\label{C1}
\end{equation}
where $w$ is a positive number.

Putting $t=(q-q_i)^{1/w}$, we evaluate each term of the expansion as
\begin{eqnarray}
\frac{1}{2\pi \imat}\int_{C}(q-q_i)^{\frac{n}{w}}dq
=  
\frac{w}{2\pi \imat}\oint t^{n+w-1}dt
= 
\left\{\begin{array}{ll}w & {\rm if} ~n+w=0  \\
0 & {\rm otherwise},
\end{array}\right.
\end{eqnarray}
where $C$ is a closed curve circling around the point $q=q_{i}$ 
$w$ times. 

For a Hamiltonian of the form $H=p^2/2+V(q)$ where $V(q)$ is a polynomial 
function of $q$, 
the function $p(q)$ does not contain negative 
order terms in the corresponding Puiseux series. 
Therefore the action integrals for 
the $\gamma_i$ loops all vanish. 
For the normal form Hamiltonian (\ref{NFgeneral}), 
if the condition $2k>m$ holds, 
the $\gamma_i$ contributions are all zero as well since $p(q)$ 
does not contain negative order terms in the Puiseux series. 
On the other hand, for $2k \le m$, the coefficient for the highest order of $p$ 
contains the variable $q$, resulting in a non-zero contribution from $\gamma_i$ loops.

\section{The action relation and the residue at infinity}
In this appendix, we provide an explicit derivation of action relations. 
The following calculations can easily be generalized to the multi-well potential systems. 
We here present double- and triple-well cases as examples. 

\subsection{Double-well Case}

\begin{figure}[htbp]
\begin{center}
\includegraphics[width=7cm]{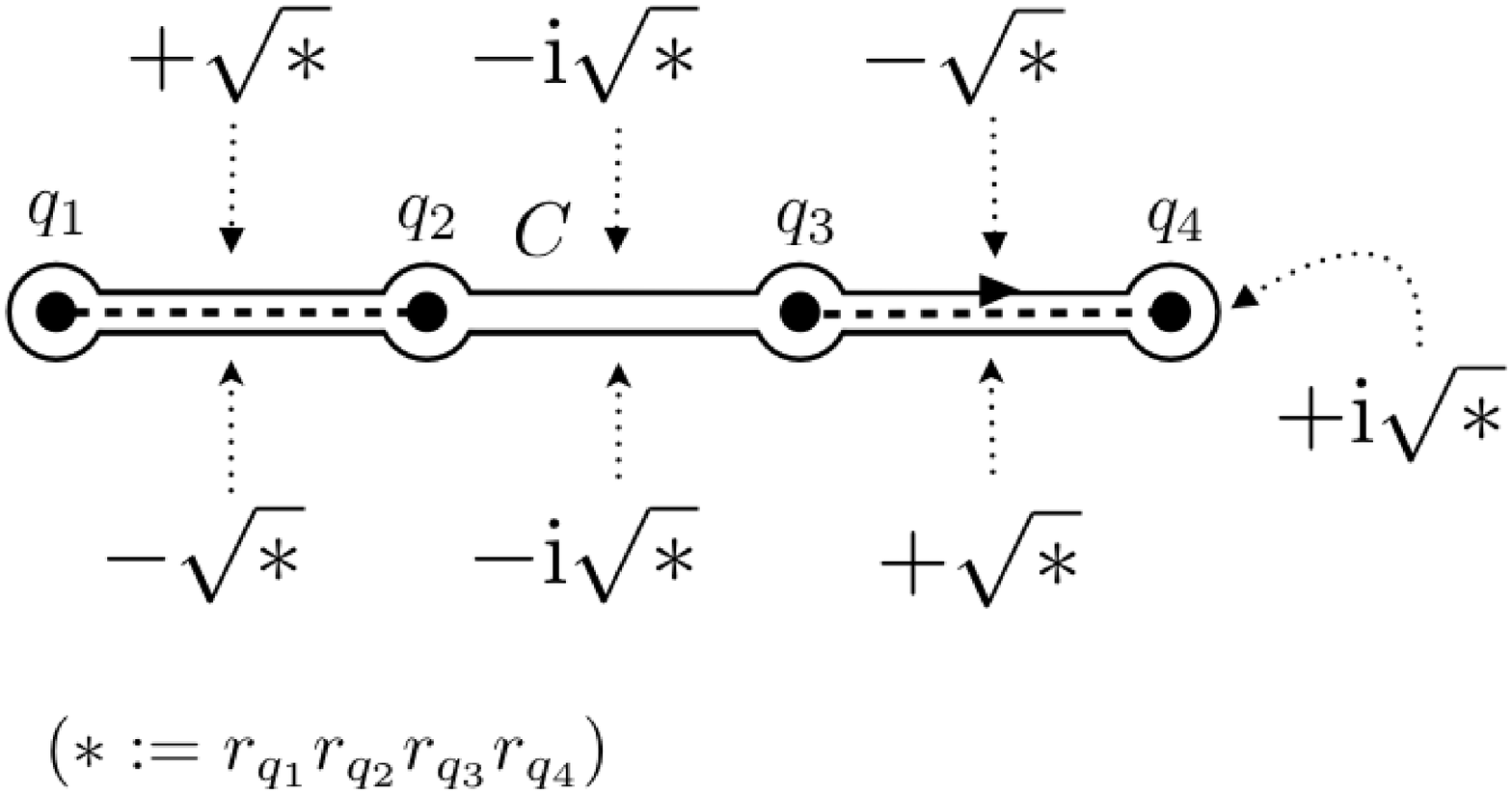}
\caption{The integration contour $C$ and the phase of $p(q)$ at each position. The black dots are branch points. The dash lines represent the branch cuts.}
\label{contour}
\end{center}
\end{figure}

Let us consider the Hamiltonian:
\begin{eqnarray}
H= \frac{p^2}{2}+V(q),
\end{eqnarray}
where $V(q)=E+(q-q_1)(q-q_2)(q-q_3)(q-q_4)$.
Branch points of $p(q)$ at the energy $E$ are located at $q=q_i ~ (1\le i \le 4)$. 
Let $C$ be a closed curve rotating clockwise around all branch points (figure \ref{contour}). 
The loop $C$ is homotopic to the loop around infinity on the Riemann sphere, 
so the integration
 along this loop is equal to the residue of infinity. 
We calculate the residue at $q=+\infty$ as follows.
Introducing a new coordinate $q=1/\eta$,
we find 
\begin{eqnarray}
\oint_{\Gamma^{(\infty)}} p(q)dq&=&\oint \frac{1}{\eta^2}\sqrt{-W(\eta)}(\frac{-1}{\eta^{2}})d\eta\nonumber\\
&=&\oint \frac{-1}{\eta^4}\left( \sum_{k}C_{k}\eta^k\right) d\eta,\nonumber
\end{eqnarray}
where $W(\eta):= 2(1-q_1\eta)(1-q_2\eta)(1-q_3\eta)(1-q_4\eta)$, 
and $C_{k} ~ (k \ge 0)$ are coefficients of the Taylor expansion of $\sqrt{-W(\eta)}$. $\Gamma^{(\infty)}$ denotes a single loop encircling $\eta=0$.
For the integration over $\eta$, the loop rotates anticlockwise around $\eta=0$, 
and the residue is evaluated as $-2\pi \imat C_{3}$. An explicit form of $C_{3}$ is 
\begin{eqnarray}
C_{3}&=&\imat \Bigl(\frac{1}{4}(q_1 + q_2 + q_3 + q_4)(q_1 q_2 + q_1 q_3 + q_2 q_3 - \frac{1}{4}(q_1 + q_2 + q_3 + q_4)^2 \nonumber\\
 &&+ q_1 q_4 + q_2 q_4 + q_3 q_4)+ \frac{1}{2} (-q_1 q_2 q_3 - q_1 q_2 q_4 - q_1 q_3 q_4 - q_2 q_3 q_4)\Bigr). \nonumber
\end{eqnarray}
Hence we obtain $S^{(\infty)}:=\oint_{\Gamma^{(\infty)}} p(q)dq=-2\pi \imat C_{3}$.

On the other hand, we evaluate the same integral by 
taking the integration along the real axis. 
We introduce new coordinates $r_{q_{i}}$ and $\theta_{q_{i}}$ as 
$r_{q_i}e^{\imat\theta_{q_i}}:=q-q_i ~(1 \le i \le 4)$. 
Here we have to take a close look at the phase of the function $p(q)$ and 
the upper limit of the integration. 
If we take the phase as $p(q)=\imat\sqrt{r_{q_1}r_{q_2}r_{q_3}r_{q_4}}$, 
the upper limit should satisfy the condition $q_4<q$ in order that 
the phase of $p(q)$ is consistent with the residue calculation at infinity, 
as shown in figure \ref{contour}. We therefore obtain 
\begin{eqnarray}
\oint_{C} p(q)dq&=&2\int_{q_1}^{q_2}pdq-2\int_{q_3}^{q_4}pdq\nonumber\\
&=&S_{L}-S_{R}.
\label{SR1}
\end{eqnarray}
Finally we get the relation (\ref{DWR})
\begin{eqnarray}
S^{(\infty)}=\oint_{\Gamma^{(\infty)}}  p(q)dq=\oint_{C} p(q)dq=S_{L}-S_{R}. 
\label{SR1}
\end{eqnarray}

\subsection{Triple-well Case}
For the triple well case where $V(q)=E+(q-q_1)(q-q_2)(q-q_3)(q-q_4)(q-q_5)(q-q_6)$, 
we find 
\begin{eqnarray}
\oint_{\Gamma^{(\infty)}} p(q)dq&=&\oint \frac{1}{\eta^3}\sqrt{-W(\eta)}(\frac{-1}{\eta^{2}})d\eta\nonumber\\
&=&\oint \frac{-1}{\eta^5}\left( \sum_{k}C_{k}\eta^k\right) d\eta. \nonumber
\end{eqnarray}
Here $W(\eta):=2(1-\eta q_1)(1-\eta q_2)(1-\eta q_3)(1-\eta q_4)(1-\eta q_5)(1-\eta q_6)$, 
and $C_{k} ~ (k \ge 0)$ are coefficients of the Taylor expansion of $\sqrt{-W(\eta)}$. 
The residue is evaluated as $-2\pi \imat C_{4}$. 
Hence we obtain $S^{(\infty)}:=\oint_{\Gamma^{(\infty)}} p(q)dq=-2\pi \imat C_{4}$.

\begin{figure}[htbp]
\begin{center}
\includegraphics[width=10cm]{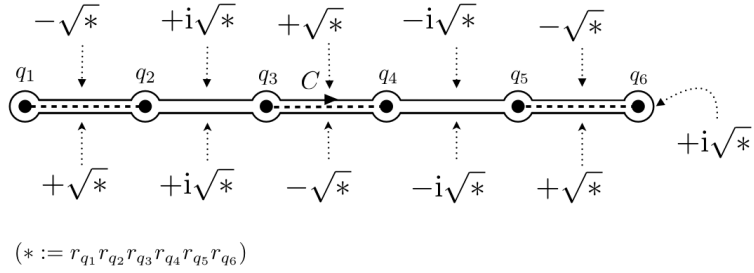}
\caption{The integration contour $C$ and the phase of $p(q)$ at each position. The black dots are branch points. The dash lines represent the branch cuts.}
\label{contour2}
\end{center}
\end{figure}

On the other hand, we calculate the same integral along the real axis. 
As shown in figure \ref{contour2}, we choose a closed curve $C$ rotating clockwise around all branch points, and 
introduce new coordinates 
$r_{q_{i}}$ and $\theta_{q_{i}}$ as 
$r_{q_i}e^{\imat\theta_{q_i}}:=q-q_i ~(1 \le i \le 6)$. 
If we take the phase as $p(q)=\imat\sqrt{r_{q_1}r_{q_2}r_{q_3}r_{q_4}r_{q_5}r_{q_6}}$, 
the upper limit should satisfy the condition $q_6<q$ in order that 
the phase of $p(q)$ should be consistent with the residue calculation at infinity, 
as shown in figure \ref{contour2}.
Then we obtain
\begin{eqnarray}
\oint_{C} p(q)dq&=&-2\int_{q_1}^{q_2}pdq+2\int_{q_3}^{q_4}pdq-2\int_{q_5}^{q_6}pdq\nonumber\\
&=&-S_{L}+S_{C}-S_{R}.
\label{SR2}
\end{eqnarray}
Finally we reach the relation (\ref{actionrerationintriple})
\begin{eqnarray}
S^{(\infty)}=-S_{L}+S_{C}-S_{R}.
\label{SR2}
\end{eqnarray}

\newpage
\printendnotes

\end{document}